\documentclass[]{achemso}

\usepackage{graphicx}
\usepackage{xcolor}
\usepackage[draft]{fixme}

\usepackage{amsmath}
\usepackage{latexsym}
\usepackage{float}
\usepackage{enumerate}
\usepackage{amsfonts}
\usepackage{amssymb}
\usepackage{color}
\usepackage{lipsum}

\title{Combining scattering experiments and colloid theory to characterize charge effects in concentrated antibody solutions}

 \author{Alessandro Gulotta}
 \affiliation{Physical Chemistry, Department of Chemistry, Lund University, SE-221 00 Lund, Sweden}

 \author{Marco Polimeni}
 \affiliation{Physical Chemistry, Department of Chemistry, Lund University, SE-221 00 Lund, Sweden}
 
  \author{Samuel Lenton}
 \affiliation{Physical Chemistry, Department of Chemistry, Lund University, SE-221 00 Lund, Sweden}
 
 \author{Charles G. Starr}
   \affiliation{Biologics Drug Product Development and Manufacturing, CMC Development, Sanofi, Framingham, MA, 01701, USA}
 
 \author{Jonathan S. Kingsbury}
  \affiliation{Biologics Drug Product Development and Manufacturing, CMC Development, Sanofi, Framingham, MA, 01701, USA}
 
 \author{Anna Stradner} 
\affiliation{Physical Chemistry, Department of Chemistry, Lund University, SE-221 00 Lund, Sweden}
\alsoaffiliation{LINXS - Lund Institute of Advanced Neutron and X-ray Science, Scheelevägen 19, SE-223 70 Lund, Sweden}

\author{Emanuela Zaccarelli} 
\affiliation{Institute for Complex Systems, National Research Council (ISC-CNR), Piazzale Aldo Moro 5, 00185 Rome, Italy}
\alsoaffiliation{Department of Physics, Sapienza University of Rome, Piazzale Aldo Moro 2, 00185 Rome, Italy}

\author{Peter Schurtenberger} \email{peter.schurtenberger@fkem1.lu.se}
\affiliation{Physical Chemistry, Department of Chemistry, Lund University, SE-221 00 Lund, Sweden}
\alsoaffiliation{LINXS - Lund Institute of Advanced Neutron and X-ray Science, Scheelevägen 19, SE-223 70 Lund, Sweden}

\begin{document}

\begin{abstract}
Charges and their contribution to protein-protein interactions are essential for the key structural and dynamic properties of monoclonal antibody (mAb) solutions. In fact, they influence the apparent molecular weight, the static structure factor, the collective diffusion coefficient or the relative viscosity and their concentration dependence. Further, charges play an important role in the colloidal stability of mAbs. There exist standard experimental tools to characterise mAb net charges such as the measurement of the electrophoretic mobility, the second virial coefficient, or the diffusion interaction parameter. However, the resulting values are difficult to be directly related to the actual overall net charge of the antibody and to theoretical predictions based on its known molecular structure. 

Here, we report the results of a systematic investigation of the solution properties of a charged IgG1 mAb as a function of concentration and ionic strength using a combination of electrophoretic measurements, static and dynamic light scattering, small-angle x-ray scattering (SAXS), and tracer particle-based microrheology. We analyse and interpret the experimental results using established colloid theory and coarse-grained computer simulations. We discuss the potential and limits of colloidal models for the description of interaction effects of charged mAbs, in particular pointing out the importance of incorporating shape and charge anisotropy when attempting to predict structural and dynamic solution properties at high concentrations.

\end{abstract}

\maketitle

\section{Introduction}
Stability against aggregation and self-assembly, low viscosity, and low opalescence at high concentrations are essential attributes required from promising high-concentration formulations of monoclonal antibodies (mAbs). Charges play a crucial role in achieving these properties \cite{Kingsbury2020}, and there are a number of studies that focused on the role of charges in mAbs \cite{Roberts2014, Yang2019, Yadav2011, Yadav2012, Hebditch2019}. A major problem in experimentally assessing mAb charge is caused by the fact that the experimental techniques such as electrophoretic measurements, static light scattering or small-angle scattering provide only so-called effective charges $Z_{eff}$ \cite{Ware1974, Miller2020, Naegele1996, Filoti2015}. With these techniques, quantities such as the electrophoretic mobility $\mu_e$ or the static structure factor $S(q)$ are experimentally measured, and the effective charge is then calculated based on explicit models, such as a non-conducting sphere with hardcore and/or homogeneous charge distribution on the surface. At the same time, there exist numerical approaches to calculate mAb charges, $Z_{cal}$, using the known molecular composition and the pKa values of the different amino acids as a function of solution conditions \cite{Johnson1994ReactiveCM, MOE}. Unfortunately, it is a common observation that $Z_{cal}$ and $Z_{eff}$ are in general very different, even when using exactly the same solvent conditions and molecular composition in experiments and simulations/calculations. Moreover, $Z_{eff}$ values determined with electrophoretic and scattering methods also disagree with each other. For mAb solutions, $Z_{eff}$ is usually experimentally determined at relatively low concentrations, i.e. in the so-called virial regime, where interactions between mAbs can either be neglected, as in the case of electrophoretic measurements, or are interpreted using virial theories, focusing on effective interaction parameters such as the second virial coefficient $B_2$ or the diffusion interaction parameter $k_D$ \cite{Roberts2014, Yang2019, Yadav2011, Yadav2012, Kingsbury2020}.

In this paper, we investigate the role of mAb charges on different structural and dynamic properties, such as the apparent molecular weight $\big\langle M_{w} \big\rangle_{\mathrm{app}}$, the static structure factor $S(q)$, the electrophoretic mobility $\mu_e$, the collective diffusion coefficient $D_c$ or apparent hydrodynamic radius $\big\langle R_{h} \big\rangle_{\mathrm{app}}$, and the relative viscosity $\eta_r$ as a function of mAb concentration and ionic strength. We use established colloid theories and assess whether they allow for a consistent description of the experimental quantities over the full range of concentrations. 

Specifically, we use a simple coarse-grained model where the mAb is described as a hard sphere interacting via an effective pair potential based on three contributions arising from the excluded volume, screened Coulomb and short-range attractive interactions.
We show that even though the obtained agreement between theoretical predictions and experimental observations is surprisingly good, for the effective charge, which is the key parameter of interest in the present work, we observe systematic differences between the values obtained from electrophoretic light scattering, static light scattering/SAXS and the theoretical charge based on the molecular composition of the mAb. We thus also discuss possible improvements in the coarse-graining strategy using either computer simulations or numerical calculations, that would allow to make more quantitative predictions of the actual solution properties based on the molecular mAb structure only. We demonstrate that computer simulations implementing a relatively simple bead model that mimics the Y-shaped anisotropic structure of the mAb indeed result in a much better agreement between $Z_{cal}$ and $Z_{eff}$, and are also able to reproduce the local structural features of the mAb solutions described by $S(q)$ from SAXS at all investigated concentrations.

\section{Experimental results}

\begin{figure}[!hbt]
\includegraphics[width=0.45\linewidth]{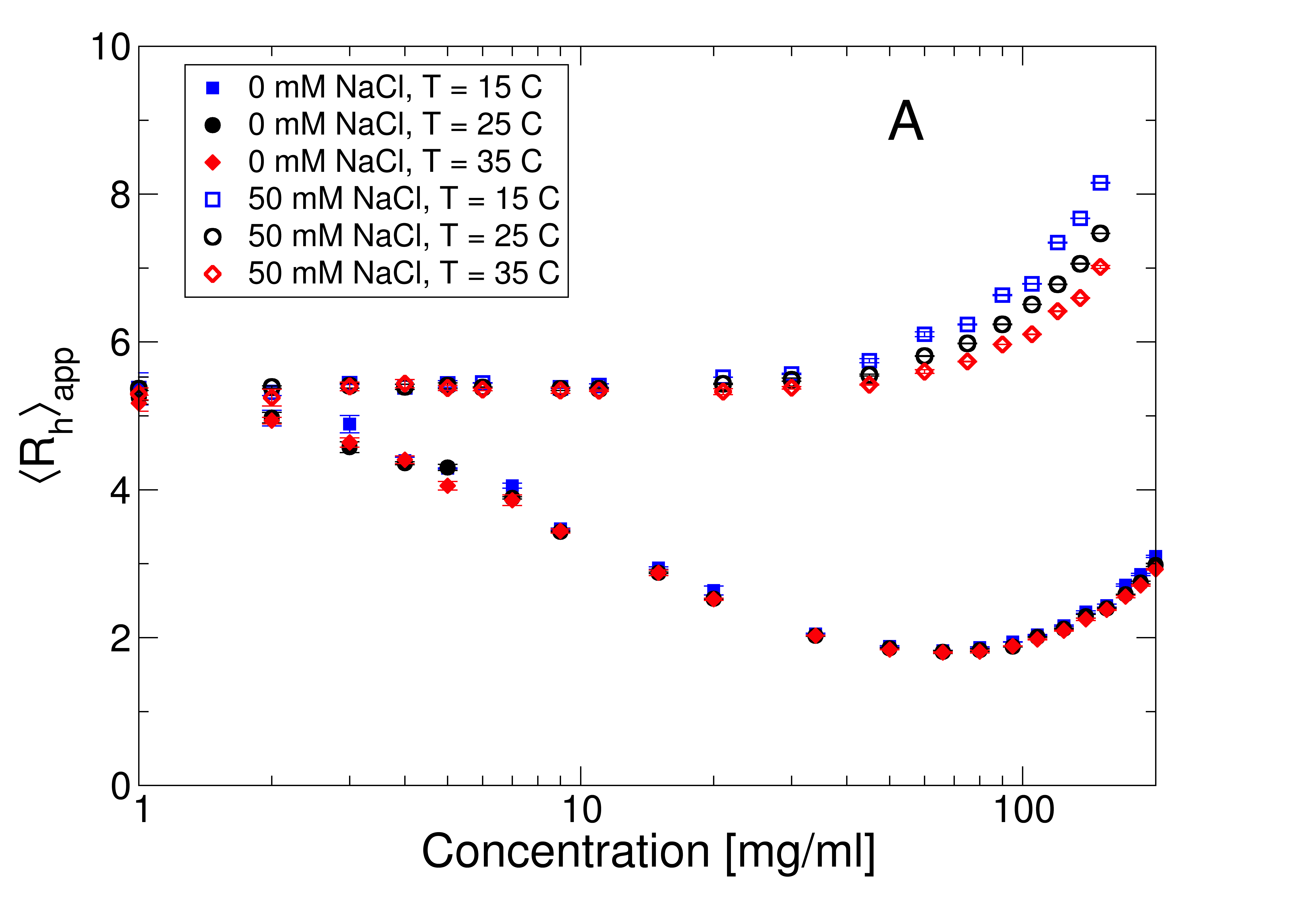}
\includegraphics[width=0.45\linewidth]{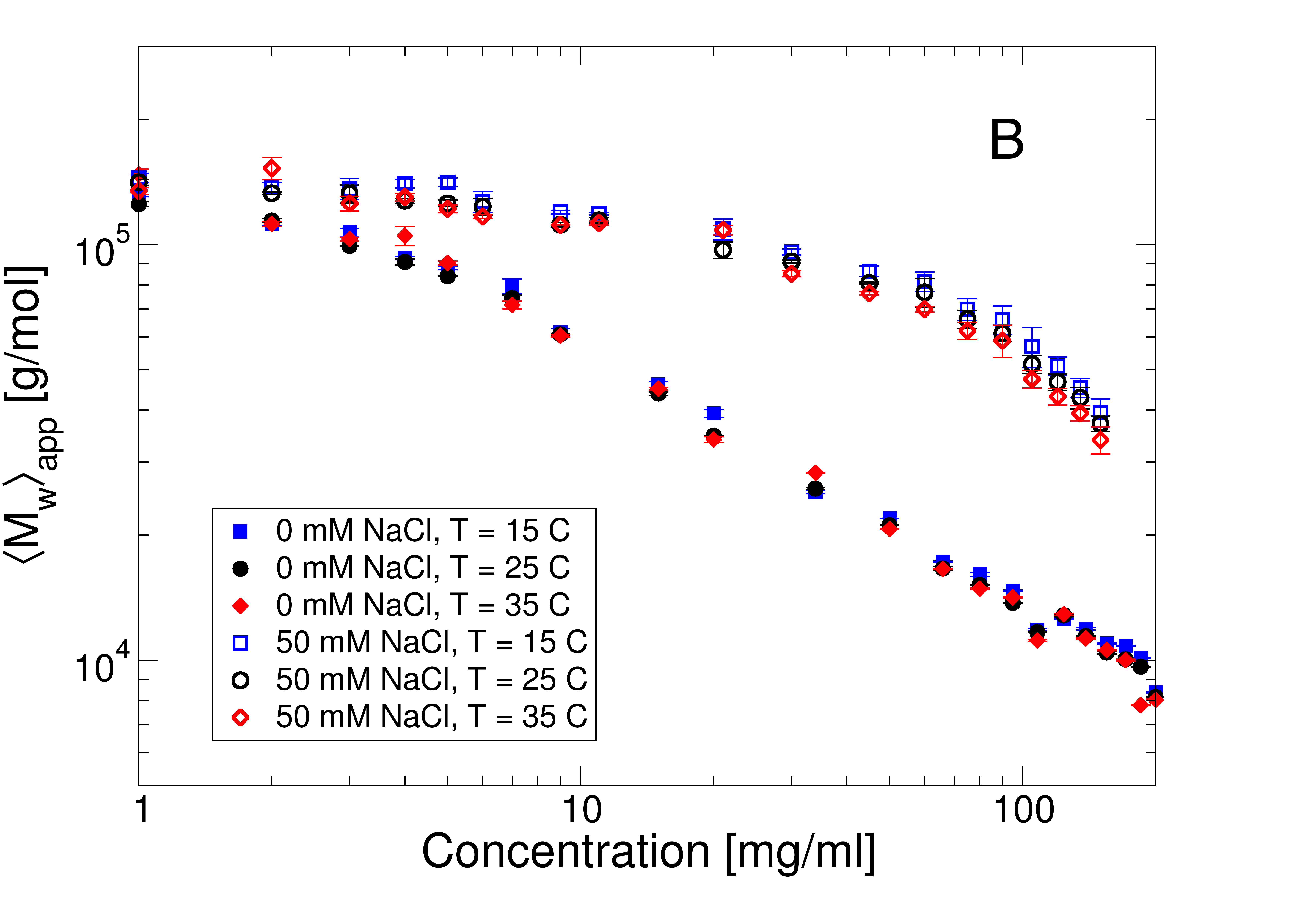}
\caption{A: $\big\langle R_{h} \big\rangle_{\mathrm{app}}$ and B:$\big\langle M_{w} \big\rangle_{\mathrm{app}}$ vs. \emph{c} 
measured as a function of mAb concentration \emph{c} at different temperatures: 15 (blue squares), 25 (black circles) and, 35 (red diamonds) $^{\circ}$C. Filled symbols correspond to no added salt, open symbols to an additional 50 mM NaCl.  }
\label{fig:SLS-DLS_mAbG}
\end{figure}

The apparent hydrodynamic radii $\big\langle R_{h} \big\rangle_{\mathrm{app}}$ of the mAb solutions are shown in Fig. \ref{fig:SLS-DLS_mAbG} as a function of concentration and temperature for two different values of the ionic strength. With no added salt, where the buffer provides an ionic strength of 7 mM, we observe a $T-$independent decrease of $\big\langle R_{h} \big\rangle_{\mathrm{app}}$ in the concentration range 1 $\leq$ \emph{c} $\leq$ 70 mg mL$^{-1}$, from $\big\langle R_{h} \big\rangle_{\mathrm{app}} \approx 5$ to $\big\langle R_{h} \big\rangle_{\mathrm{app}} \approx 2$ nm. This is typical for a repulsive system where the interactions are likely dominated by a combination of screened Coulomb repulsion due to the low ionic strength and excluded volume effects. For c $\geq$ 70 mg mL$^{-1}$, the apparent hydrodynamic radius shows a dramatic increase. Upon the addition of 50 mM NaCl, i.e. with a total ionic strength of 57 mM, the overall behaviour changes quite dramatically. $\big\langle R_{h} \big\rangle_{\mathrm{app}}$ initially remains constant, indicating that excluded volume interactions are now compensated by an additional attractive interaction. For c $\geq$ 50 mg mL$^{-1}$, the apparent hydrodynamic radius again increases strongly, and we now also observe a clear temperature dependence. \\

The apparent molecular weight $\big\langle M_{w} \big\rangle_{\mathrm{app}}$  obtained by static light scattering is also shown in Fig. \ref{fig:SLS-DLS_mAbG} at $T=$ 15, 25 and 35 $^{\circ}$C and two ionic strength values, respectively. With no added salt, there is no measurable $T$ dependence. In contrast to  $\big\langle R_{h} \big\rangle_{\mathrm{app}}$, which was found to increase at high concentrations, the $\big\langle M_{w} \big\rangle_{\mathrm{app}}$ data decrease monotonically at low ionic strength, again indicating a purely repulsive behaviour, with no sign of formation of larger aggregates at higher concentrations. For the case with added 50 mM NaCl, $\big\langle M_{w} \big\rangle_{\mathrm{app}}$ also decreases monotonically with increasing concentration for all three temperatures, but there is now a small but systematic dependence on temperature, and the repulsive interactions appear to be significantly smaller. 
\\

\begin{figure}[!hbt]
\includegraphics[width=0.8\linewidth]{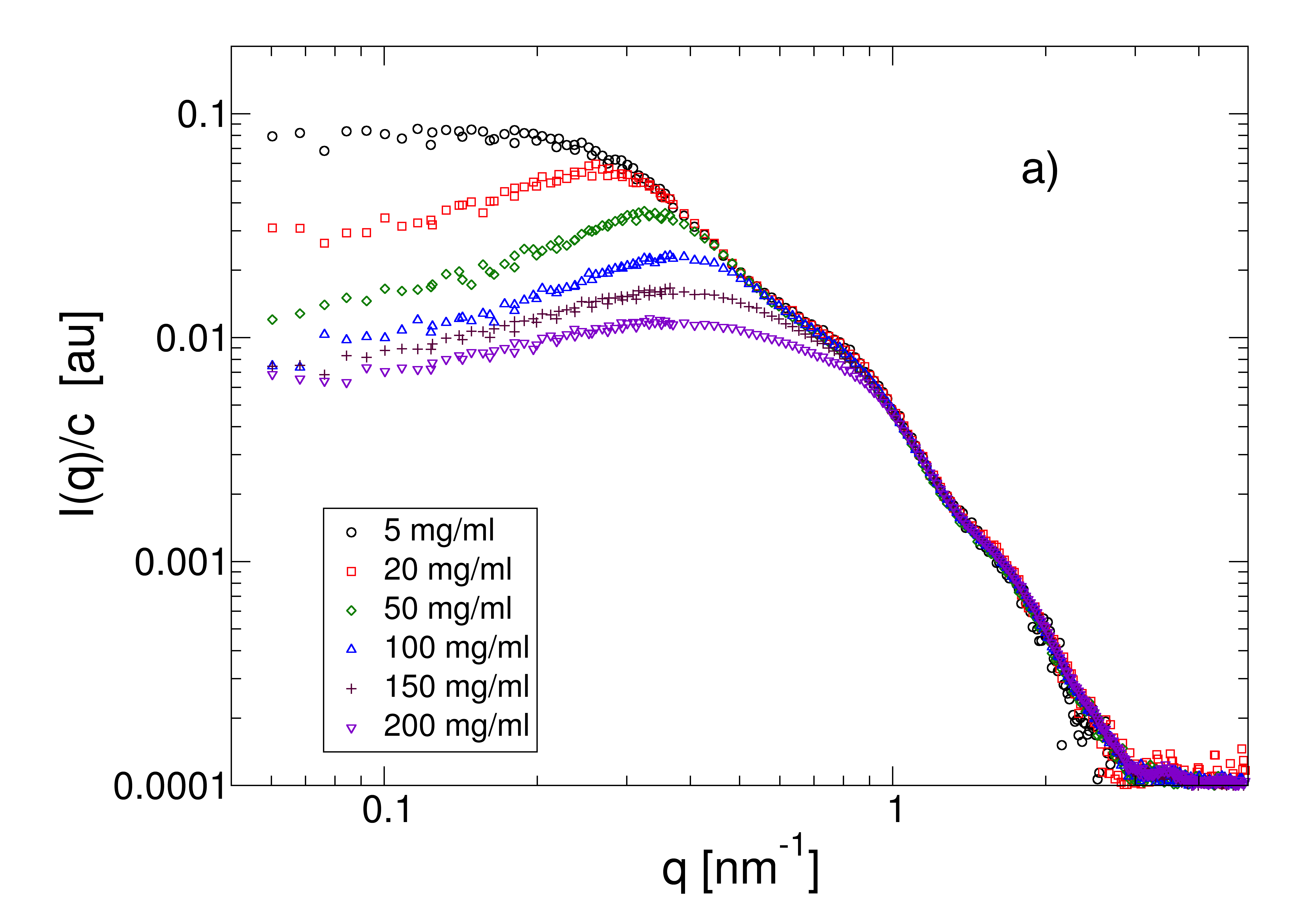}
\includegraphics[width=0.8\linewidth]{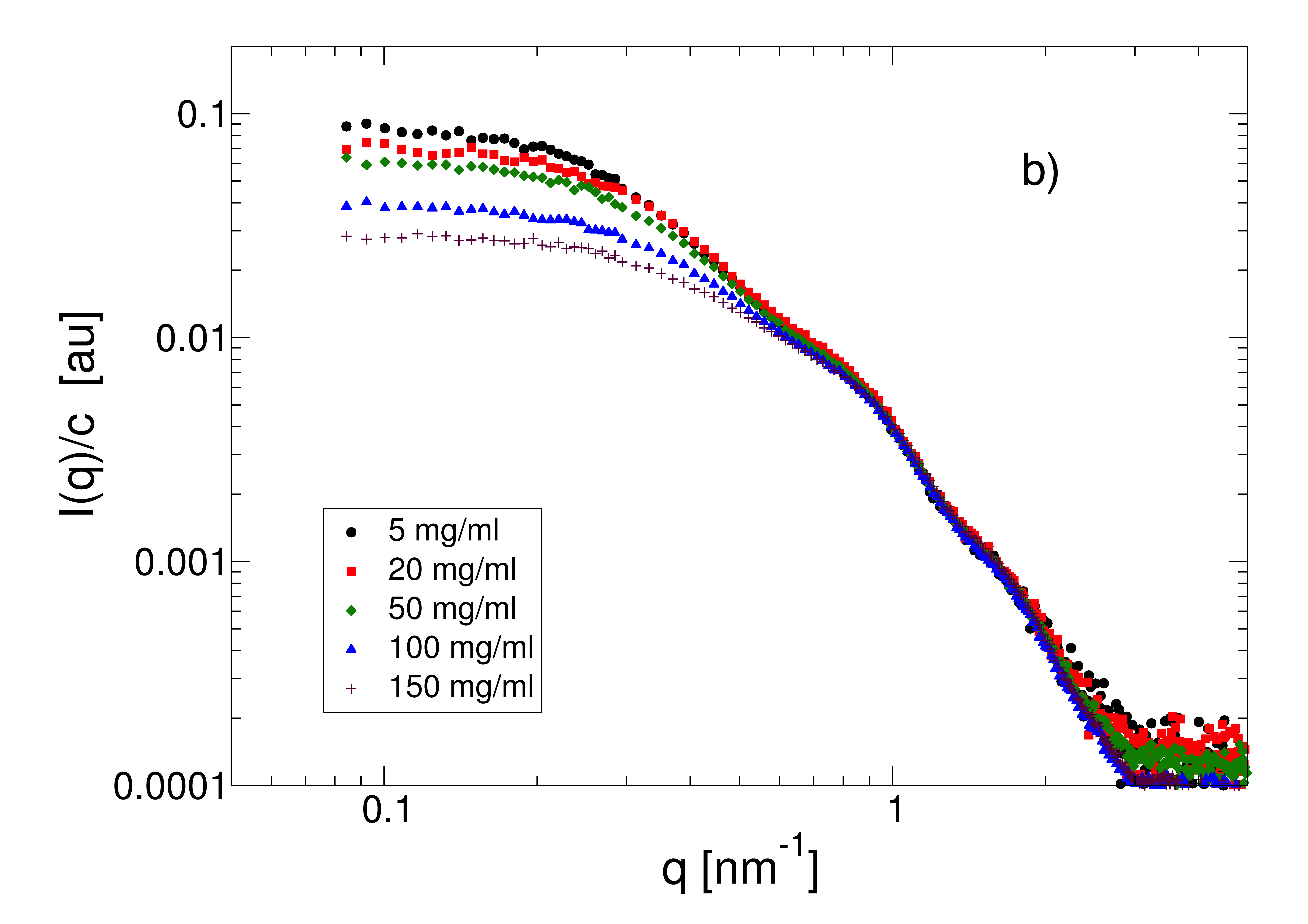}
\caption{Concentration-normalized small-angle scattering data $I(q)/c$ vs. \emph{c} for 25 $^{\circ}$C and no added salt (a) or with added 50 mM NaCl (b). Actual concentrations are given in the legend box of the two graphs. }
\label{fig:Iq}
\end{figure}

Additional high-resolution information about the solution structure can be obtained with SAXS. In Fig. \ref{fig:Iq} we summarize the data obtained at $T=$ 25 $^{\circ}$C for different mAb concentrations. Fig. \ref{fig:Iq}(a) shows the data with no added salt. We see a strong decrease of the scattering data at low $q$-values with increasing concentration, analogous to the $c$-dependence of $\big\langle M_{w} \big\rangle_{\mathrm{app}}$  measured by static light scattering, and we also observe an indication of a weak structure factor peak at $q$-values around 0.3 - 0.4 nm$^{-1}$, while the scattering data at higher values of $q$ all overlap for the different studied concentrations. This clearly indicates that the solution structure of the mAb solutions is dominated by repulsive interactions that lead to increasingly strong positional correlations, but that concentration has no measurable effect on the mAb structure. The data obtained with 50 mM NaCl added, reported in Fig. \ref{fig:Iq}(b), show significantly weaker interaction effects with increasing concentration, in line with the results from static and dynamic light scattering. Structural correlations appear to be much less pronounced due to the strongly screened electrostatic interactions. However, once again the high $q$-data overlap, although the scatter of the points at lower concentrations is larger. This is due to the decreased scattering contrast of the mAb against the solvent, because of the added salt.

\begin{figure}[!hbt]
\centerline{\includegraphics[width=0.8\linewidth]{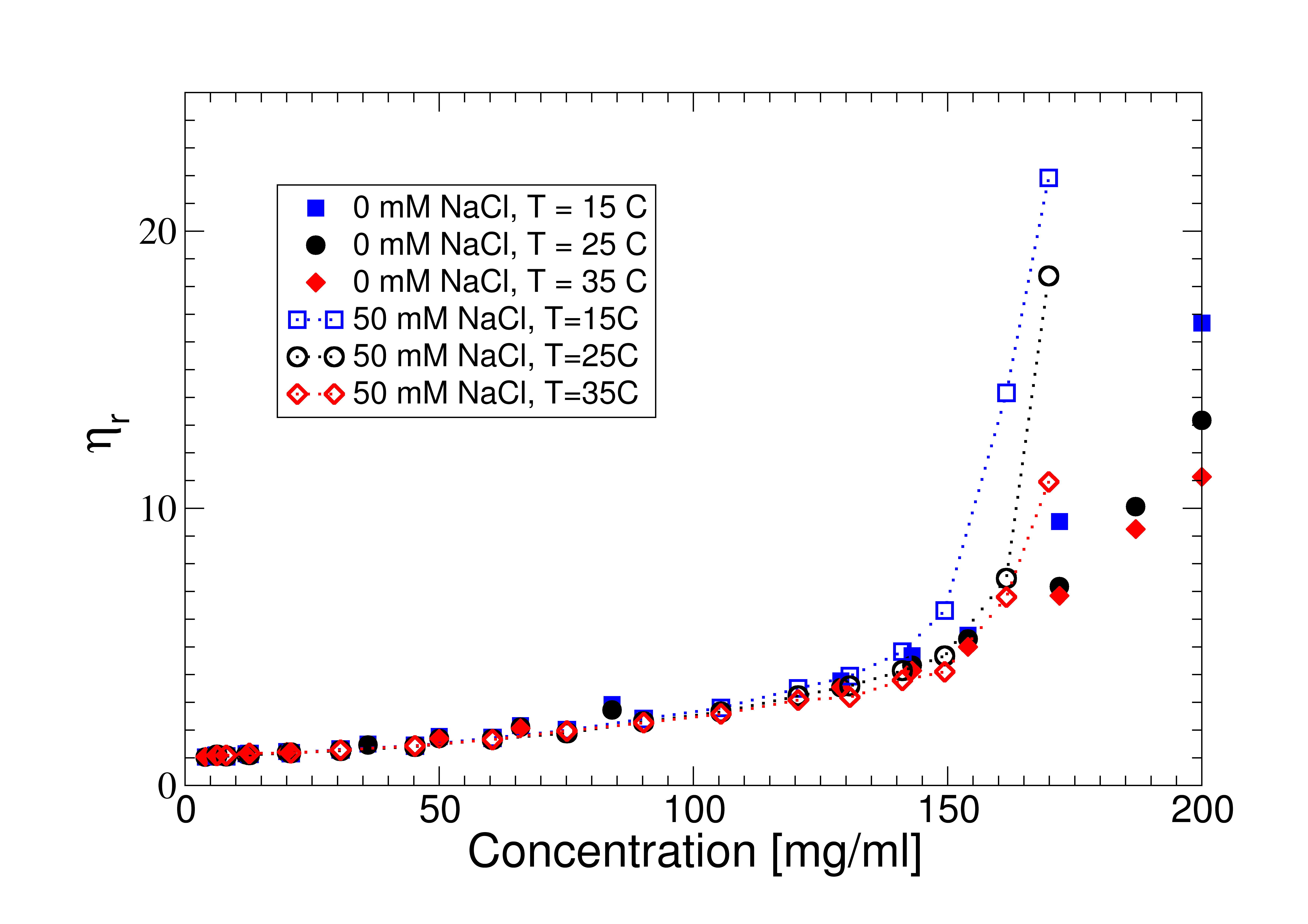}}
\caption{Relative viscosity $\eta_r$ vs. \emph{c} for 15 $^{\circ}$C (solid blue symbols), 25 $^{\circ}$C (solid black symbols) and 35 $^{\circ}$C (solid red symbols),  with no added salt, and with 50 mM NaCl added for 15 $^{\circ}$C (open blue symbols), 25 $^{\circ}$C (open black symbols) and 35 $^{\circ}$C (open red symbols), respectively. }
\label{fig:etar-exp}
\end{figure}

The results from measurements of the relative viscosity $\eta_r = \eta_0/\eta_s$, where $ \eta_0$ is the zero shear viscosity of the mAb solution and  $\eta_s$ the solvent viscosity, respectively, are shown in Fig. \ref{fig:etar-exp}. For concentrations smaller than about 120 mg/ml we see no significant influence of either temperature nor ionic strength, and $\eta_r$ increases weakly with increasing $c$. However, at higher concentrations, the different solvent conditions have a dramatic effect on the relative viscosity. For low ionic strength, $\eta_r$ exhibits a behaviour that is typical for mAb solutions with weak self-assembly, where the increase of the viscosity appears to be most pronounced for the lowest temperature, in agreement with observations for other globular protein systems that undergo equilibrium cluster formation \cite{Stradner2004, Stradner2006b, Cardinaux2011, Bergman2019}. For the higher ionic strength, the effect of concentration is much more dramatic, and the viscosity appears to diverge at a much lower protein concentration, and with significantly different qualitative behaviour.

\section{Discussion}

\subsection{Structural properties}
We first attempt to analyse and understand the static properties of mAb solutions as characterised by SLS and SAXS. The scattering intensity $I(q)$ measured in these experiments is related to the static structure factor $S(q)$, \cite{Schurtenberger1993}
\begin{equation}
	\label{intensity}
	I(q) = A M_w c P(q) S(q)
\end{equation}
\noindent where $A$ is a constant that combines instrument parameters and contrast terms, $P(q)$ is the particle form factor, and $S(q)$ is the structure factor. In the case of polydisperse particles and/or anisotropic particle shape, $P(q)$ and $S(q)$ are so-called effective or measured quantities \cite{klein2002interacting, Greene2016}. \\

\begin{figure}[!h]
\includegraphics[width=1\linewidth]{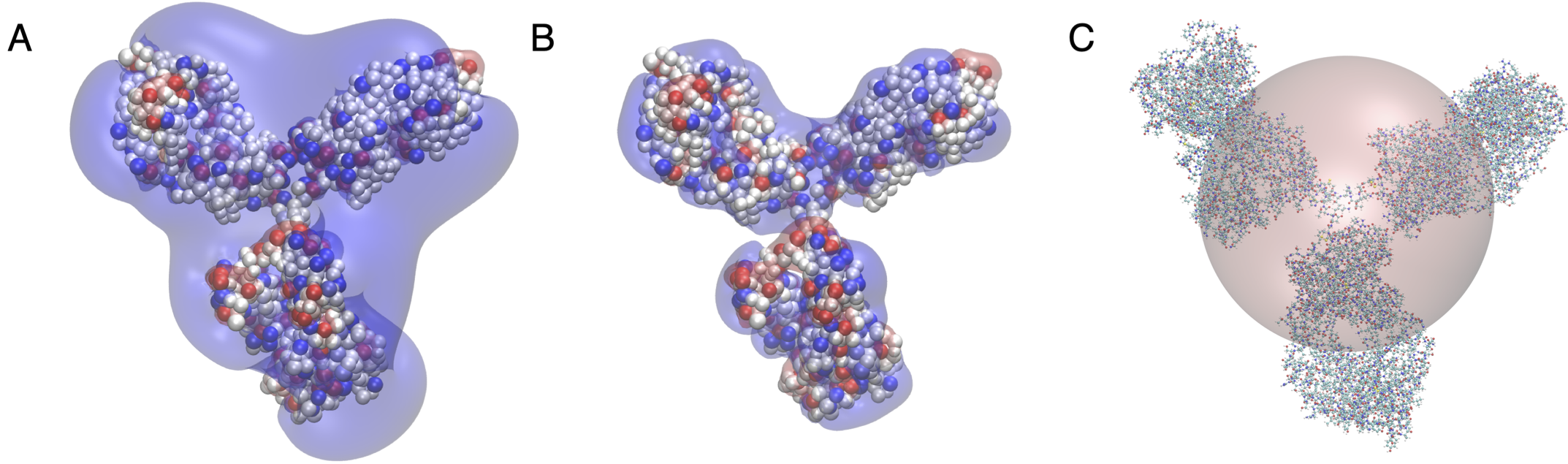}
\caption{Differently coarse-grained representations of the Y-shaped antibody, consisting of two arms (so-called Fab domains) and one leg (so-called Fc domain). A, B: Images of the charge distribution and resulting isostatic potential surface. Here, we show the antibody in a slightly coarse-grained representation where each amino acid is represented by a small bead, and where coloured beads represent charged amino acids with blue corresponding to positive (+1e) and red to negative charges (-1e), superimposed by the resulting isosurfaces of the -1 $k_BT$ (red) and +1 $k_BT$ (blue) electrostatic potential calculated with the APBS (Advanced Poisson-Boltzmann Solver) tool \cite{Jurrus2018}. Shown are images for no added salt (A) and 50 mM NaCl added (B). C: Schematic representation of an effective hard sphere model of the monoclonal antibody, together with its all-atom representation. 
}
\label{CoarseGrainedModel}
\end{figure}

Any attempt to reproduce and/or interpret the measured structure factor of these solutions requires the choice of an appropriate model. In the current study, we will focus on a simple colloid model, which builds on the mAb charge calculations, as obtained from MC computer simulations, and on the resulting electrostatic potential, that is illustrated with a plot of the electrostatic isosurface potential also shown in Fig. \ref{CoarseGrainedModel}. While the mAb has a heterogeneous charge distribution with positive and negative charges, the resulting electrostatic potential is dominated by positive charges, so that other mAbs experience a rather globular +1 $k_BT$ isopotential surface that extends beyond the actual protein structure upon approach.  In our analysis, we, therefore, start with a simple coarse-grained colloid model based on hard spheres, as shown schematically in Fig. \ref{CoarseGrainedModel}A with an effective hard sphere radius $R_{hs}$, interacting with an effective potential also including a screened Coulomb or Yukawa interaction, caused by the weakly screened charges on the mAb\cite{Banchio2008, Naegele1996, Neal1984, Beresford-Smith1985}.

\begin{figure}[h]
\includegraphics[width=0.9\linewidth]{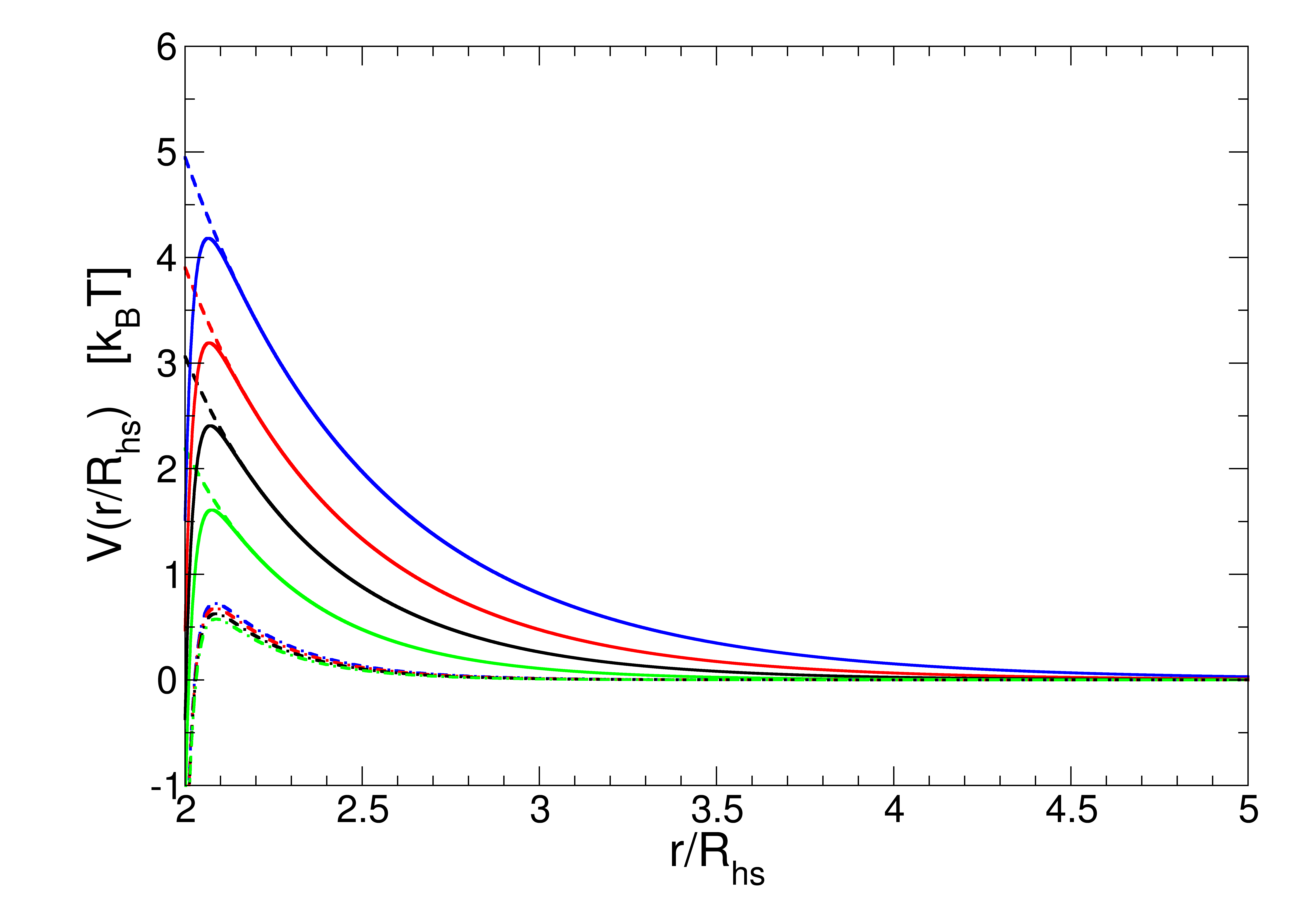}
\caption{Effective pair potential $V(r/R_{hs})$ as a function of the reduced centre-centre distance $r/R_{hs}$ for different mAb concentrations $c = 3$ mg/ml (blue lines), $c = 50$ mg/ml (red lines), $c = 100$ mg/ml (black lines) and $c = 150$ mg/ml (green lines). For an ionic strength of 7 mM (0 mM NaCl added), $V(r/R_{hs})$ is calculated using either Eq.~\ref{eq:Yukawa} (no attraction, $\epsilon_a = 0 k_BT$, dashed lines) or Eq.~\ref{eq:Pot-total} ($\epsilon_a = 3.5 k_BT$, solid lines) with $Z_{eff} = 20$ and $R_{hs} = 5$ nm. Also shown are data for  an ionic strength 57 mM (50 mM NaCl added) with $\epsilon_a = 3.5 k_BT$ (dashed-dotted lines).}
\label{Potential}
\end{figure}

The interaction potential $V(r)$ between two mAbs in this simple model can thus be written as,
\begin{equation}
\begin{aligned}
V_{\textrm{r}}(r)/k_BT = \ & \ \infty, &(r < \sigma_{\textrm{hs}})\\
=\ & \ L_B Z_{eff}^2 \bigg(\frac{e^{\kappa\sigma_{\textrm{hs}/2}}}{1 + \kappa \sigma_{\textrm{hs}}/2}\bigg)^2
 \frac{e^{-\kappa r}}{r},  &(r> \sigma_{\textrm{hs}})
\label{eq:Yukawa}
\end{aligned}
\end{equation}
\noindent with $\sigma_{\textrm{hs}} = 2 R_{hs} $ the hard sphere diameter, $Z_{\textrm{eff}}$ the effective charge of the particle, $\kappa$ the inverse Debye screening length and $L_{\textrm{B}}$ the Bjerrum length, defined as $L_{\textrm{B}} = e^2 /\epsilon_r  k_{\textrm{B}} T = 0.714$ nm (at 25 $^\circ$C). Here, $e$ corresponds to the elementary charge of one electron, $\epsilon_r$ denotes the relative dielectric constant of water, $k_{\textrm{B}}$ stands for the Boltzmann constant and $T$ is temperature. The Debye length describes the screening of the macroion charge by all microions, i.e. it includes contributions from dissociated counterions, salt, and dissociated buffer. For monovalent salt and buffer ions, it can be written as,
\begin{equation}
	\label{kappa}
	\kappa^2 =  4  \pi L_B [ (\frac{1}{1-\phi})|Z_{eff}| \rho + 2 \rho_s + 2\rho_b]
\end{equation}
\noindent where $\rho$ is the number density of particles and $\rho_s$ and $\rho_b$ are the number densities of the salt and dissociated buffer ions, respectively. The factor $1/(1 - \phi)$ corrects for the volume occupied by the proteins and thus takes into account the free volume accessible to the dissociated counterions, which cannot penetrate the protein, while we ignore the small free volume corrections arising from the finite size of the microions. The contributions from the dissociated counterions ($|Z_{eff}| \rho$) and the volume term $1/(1 - \phi)$ make the screening length, and thus the effective pair potential, concentration-dependent \cite{Beresford-Smith1985}. This is illustrated in Fig. \ref{Potential}, where $V(r)$ calculated by Eq.~\ref{eq:2} is reported for different $c$. 

To calculate an effective hard sphere volume fraction, $\phi_{hs}$, we first need to convert the experimental weight concentration into number density.    
Considering that the mass of a mAb molecule is 148 kDa, at a weight concentration of 1 mg/ml we thus have $4.068 \times 10^{15}$ particles per ml. 
$\phi_{hs}$ is then obtained by multiplying $\rho$ with the excluded volume of a single particle 
\begin{equation}
	\label{effhsvolume}
	\phi_{hs} = \rho \frac{\pi \sigma^3_{hs}}{6}
\end{equation}
\noindent where $\sigma_{hs}$ is the effective hard sphere diameter. We use a hard sphere diameter $\sigma_{hs} =2R_g = 10$ nm in Eq.~\ref{effhsvolume}, roughly equal to twice the radius of gyration of the mAb, to calculate the corresponding effective hard sphere volume fraction for a given value of $c$. 

Based on the potential in Eq.~\ref{eq:Yukawa}, we can now calculate the structure factors $S(q)$  using liquid state theory and in particular, integral equations~\cite{Goodstein1975}. The starting point is the link between the static structure factor and the pair distribution function $g(r)$ given by 
 \begin{equation}
	\label{pair-distribution}
	S(q) = 1 + 4 \pi \rho   \int_{0}^{\infty} r^2(g(r) - 1) \frac{\sin qr}{qr} \,dr\
\end{equation}
\noindent where $g(r)$ 
is calculated through an appropriate closure relation, such as the hypernetted chain (HNC) or the  Rogers-Young (RY) one\cite{klein2002interacting, Goodstein1975, Naegele1996}. In order to perform these calculations, we need to carefully choose the effective charge $Z_{eff}$, and consequently the screening constant $\kappa$ (which is related to $Z_{eff}$ through eqn. \ref{kappa}).
For the effective charge, we expect values in the range $30 \lesssim Z_{eff} \lesssim 38$. This range is based on the known molecular structure as well as on various numerical approaches based on the software Molecular Operating Environment (MOE) \cite{MOE} or from Monte Carlo simulations with the available molecular structure \cite{Polimeni2023}. In the latter, the mAb is coarse-grained at the amino acid level in order to estimate the mAb charge distribution \cite{Johnson1994ReactiveCM}. The results from different numerical procedures as well as those obtained from the analysis of our experiments described in detail below are summarised in Table~\ref{tab:charges}.

\begin{table}[h!]
\centering
\begin{tabular}{| c | c | } 
 \hline
  $ Z_{calc}$ MOE & 36.7  \\ 
 \hline
 $Z_{calc}$ MC & 31  \\
  \hline
 $Z_{eff}^\zeta$ & 12.8  \\  
 \hline
 $Z_{eff}^{RY}$ & 20  \\ 
 \hline
 $Z_{eff}^{9-bead}$ & 28  \\ [1ex] 
 \hline
\end{tabular}
\caption{Theoretical net charges and measured effective charges based on different approaches for the overall ionic strength of 7 mM: $Z_{calc}$ MOE is the net charge calculated with the software Molecular Operating Environment (MOE) using the available molecular structure; $Z_{calc}$ MC is the net charge obtained from Monte Carlo simulations at the amino acid level \cite{Polimeni2023}; $Z_{eff}^\zeta$ is the effective charge obtained from electrophoretic mobility measurements using Eq. \ref{mobility}; $Z_{eff}^{RY}$ is the effective charge obtained from the application of a colloid model based on the potential in  Eq.~\ref{eq:2} with the Rogers Young closure as compared to the SLS and SAXS data; $Z_{eff}^{9-bead}$ is the effective charge obtained from MC simulations with a 9-bead hard Y model and the comparison with the full SAXS structure factors.}
\label{tab:charges}
\end{table}


We first calculate $S(q)$ for low-concentration samples without added salt. Here we expect that the weakly screened Coulomb repulsion between the mAbs is sufficiently long-ranged and strong so that the highly coarse-grained model (Fig.  \ref{CoarseGrainedModel}c) and its associated simple effective pair potential (Fig. \ref{Potential}) should describe the real system quite well. In this case, the molecular details such as the non-spherical shape and the actual charge distribution should therefore only play a minor role.

\begin{figure}[h]
\centerline{\includegraphics[width=0.8\linewidth]{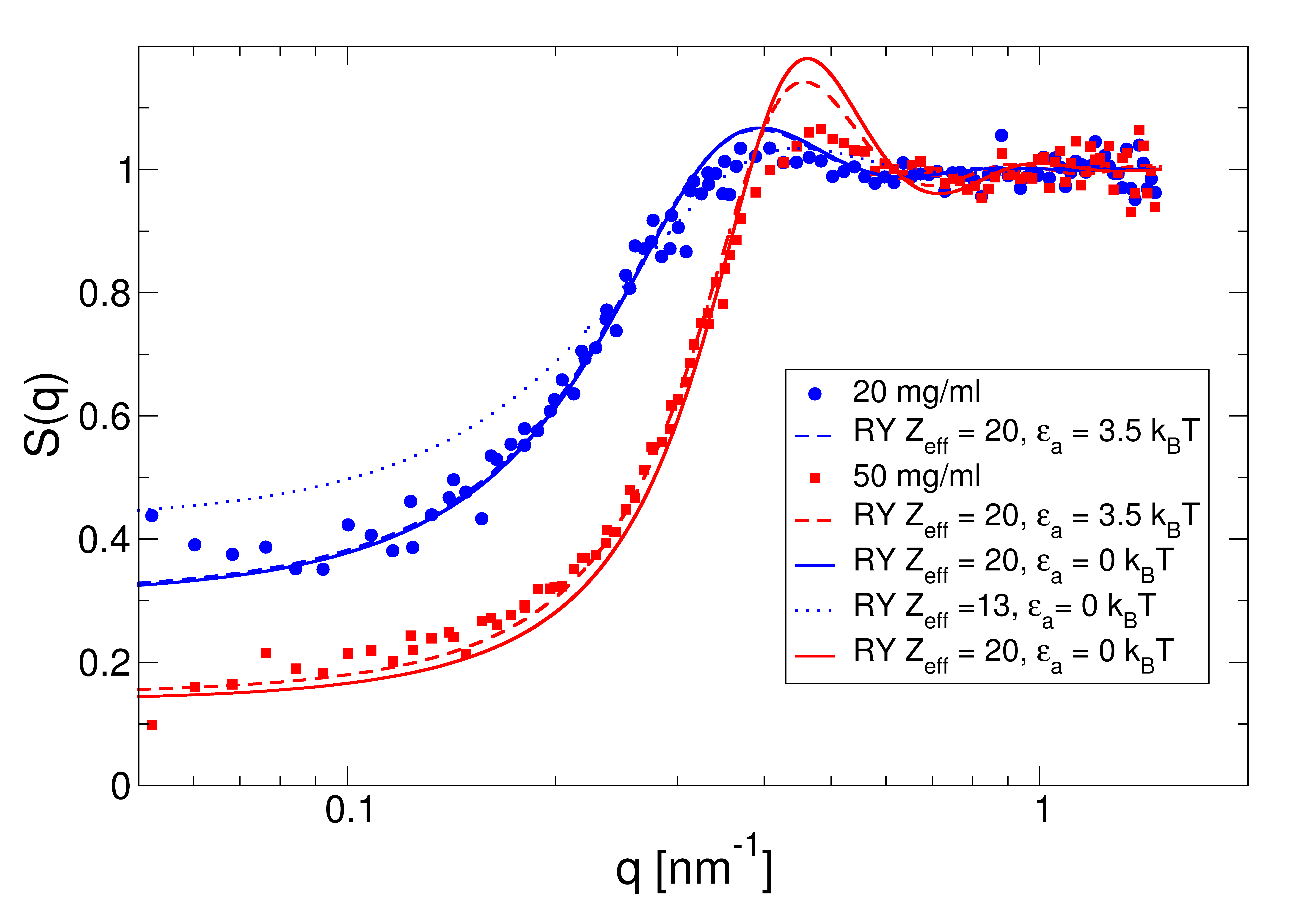}}
\caption{Experimentally determined $S(q)$ vs. \emph{q} for 25 $^{\circ}$C with no added salt compared to predictions from integral equation theory obtained through the RY closure based on the interaction potential in Eq.~\ref{eq:2}. Shown are experimental data for $c = 20$ mg/ml (blue circles) and 50 mg/ml (red squares) and theoretical curves for $c = 20$ mg/ml and $Z_{eff} = 13$ (blue dotted line), $Z_{eff} = 20$ (blue solid line) and $c = 50$ mg/ml and $Z_{eff} = 20$ (red solid line).  Also shown are calculations for a mixed potential with an additional short-range attraction of - 3.5 $k_BT$ (20 mg/ml: blue dashed line, 50 mg/ml: red dashed line).}
\label{Sq-dil-nosalt}
\end{figure}

The resulting experimental and calculated $S(q)$ within the RY closure are shown in Fig. \ref{Sq-dil-nosalt} for samples with 20 mg/ml and 50 mg/ml and no added salt, respectively. We obtain a very good agreement with $Z_{eff}^{RY} = 20$. The only systematic discrepancy between the calculations and the measured data is found in the amplitude of the nearest neighbor peak in $S(q)$, which appears more pronounced in the theoretical rather than the measured curves. This likely reflects the oversimplified structural model of perfect spheres, which becomes more important at higher concentrations, where the electrostatic potential is more strongly screened. For a Y-shaped particle, direct contact is possible for a range of interparticle distances, quite in contrast to the situation of spheres, where there is a single direct contact distance given by the particle diameter. While we, therefore, expect that the simple centrosymmetric potential shown in Fig.~\ref{Potential} should represent the actual effective pair potential between charged mAbs at low ionic strength and protein concentrations quite well, this will no longer be the case at higher ionic strength and/or high protein concentrations. Under these conditions, the additional screening from the counterions and added salt ions will result in potential values that will be low enough at larger distances to allow the mAbs to explore also smaller interparticle distances and come into direct contact. The hard sphere contribution thus becomes more important, and the non-spherical shape will then make the potential anisotropic. Ensemble-averaged pair correlation functions and structure factors will then likely show broader nearest neighbor peaks with lower amplitudes, as for example also observed in hard ellipsoids when compared to hard spheres \cite{Greene2016}.

However, we note that the predicted effective charge $Z_{eff}^{RY} = 20$ is found to be significantly below the range of net charges estimated with the different theoretical approaches discussed above.
We can also compare these results with the effective charge obtained from electrophoretic light scattering (ELS) experiments, as described in Materials and Methods, as a function of ionic strength. ELS experiments indicate that the mAb has an effective charge of around $Z_{eff}^\zeta \approx +13$, which seems independent of ionic strength up to 57 mM. The slight decrease seen at higher ionic strength could come from some ion (Cl) adsorption often seen with proteins, but systematic errors for ELS measurements at higher salt concentrations may also play a role.  Using such a low value of effective charge, the structural correlations for 20 mg/ml are clearly underestimated as also shown in Fig.  \ref{Sq-dil-nosalt}. Systematic differences between $Z_{eff}^\zeta$ and $Z_{calc}$ for mAbs were also reported previously, and primarily associated with anion binding \cite{Laber2022}.
However, it is important to realise that while ELS is often used to obtain an experimental estimate of the effective charge $Z_{eff}^\zeta$, this value results from measurements of an electrokinetic property, i.e. the electrophoretic mobility, and $Z_{eff}^\zeta$ is then calculated based on the assumption that the mAb is described by a model of a nonconducting spherical particle with a smooth and impenetrable surface and frictional properties given by the measured hydrodynamic radius extrapolated to infinite dilution, $R_h = 5.4$ nm, of the mAb. On the other hand, $Z_{eff}^{RY} = 20$ is obtained from a measurement of the structural correlations between mAbs given by the structure factor $S(q)$, i.e. based on a static property, which we calculate based on the model illustrated in Fig. \ref{CoarseGrainedModel}, with the key parameters $R_{hs}$, $Z_{eff}^{RY}$ and $\kappa$.

\begin{figure}[h]
\centerline{\includegraphics[width=0.9\linewidth]{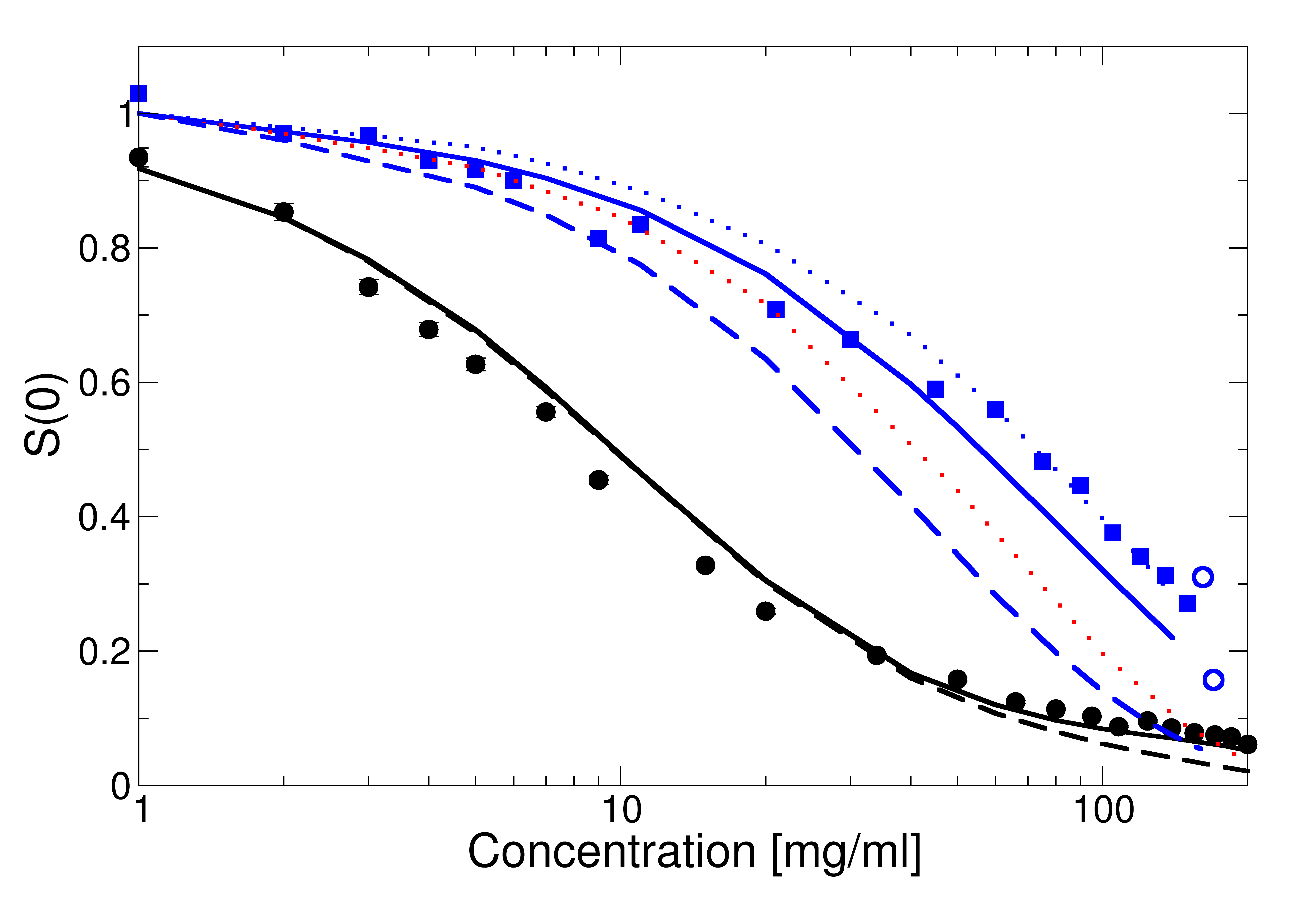}}
\caption{$S(0)$ vs. \emph{c} at 25 $^{\circ}$C with no added salt (black symbols) and an additional 50 mM NaCl (blue symbols), compared to predictions using an interaction potential based on screened Coulomb and excluded volume interactions only (Eqs. \ref{eq:Yukawa}, \ref{kappa}) using integral equation theory based on the RY closure. The dashed lines are for $Z_{eff}^{RY} = 20$ at no added salt (black) and with 50 mM added NaCl (blue), respectively. Also shown is the theoretical result for hard spheres based on the approximation by Carnahan and Starling as the red dotted line, and for an additional short-ranged attractive interaction with strength $\epsilon_a = -3.5 k_BT$ (black and blue solid lines) and for $\epsilon_a = -3.8 k_BT$ (blue dotted line). The results from the cluster model for solutions with 50 mM NaCl added are shown as open blue circles.}
\label{fig:mAb-G-S0-T25}
\end{figure}

Next, we attempt to reproduce the full concentration dependence of the SLS data for both ionic strengths. Figure~\ref{fig:mAb-G-S0-T25} compares the experimentally obtained values of the low-$q$ limit of the static structure factor, S(0), with theoretical predictions based on the charged sphere model as a function of $c$. Here, we again use the RY closure for $Z_{eff}^{RY} = 20$, where the black dashed line corresponds to no added salt, and the blue dashed line to 50 mM added NaCl, respectively. We see that the calculated $c$-dependence reproduces well the experimental data for the lower ionic strength up to concentrations around 50 mg/ml, and then appears to overestimate interaction effects at higher concentrations. It is interesting to note that the systematic deviation between the experimental and theoretical data appears at concentrations where the DLS measurements show an upturn in the concentration dependence of $\big\langle R_{h} \big\rangle_{\mathrm{app}}$ (see Fig. \ref{fig:SLS-DLS_mAbG}). It is, of course, important to realise that $Z_{eff}$ is an effective charge that for highly charged particles normally also depends on concentration \cite{Heinen2012, Naegele1996}. While this would lead to a less steep slope of $S(0)$ vs. concentration, we also see from Fig. \ref{fig:mAb-G-S0-T25} that the experimental data even crosses the hard-sphere limit at the highest concentrations, indicating that there must be a weak, but non-negligible contribution from attractive interactions.

This becomes even more clear when looking at the SLS data for solutions with increased ionic strength, i.e. with 50 mM added NaCl (Fig. \ref{fig:mAb-G-S0-T25}). While the experimental data are well reproduced by RY at the lowest concentrations $c \leq$ 10 mg/ml, at higher values of $c$ the data lie well above the theoretical values for $S(0)$, given for either charged or hard spheres. There is thus an obvious need to include an additional attractive term in the interaction potential.

To this aim we consider an additional short-range attraction, using an approach that has resulted in a quantitative description of the structural properties of concentrated solutions of globular proteins such as lysozyme that form equilibrium clusters at low ionic strength \cite{Cardinaux2007, Cardinaux2011}. The total effective pair potential $V_{\textrm{t}}(r)$ now reads as the sum of a repulsive ($V_{\textrm{r}}(r)$) and an attractive ($V_{\textrm{a}}(r)$)) term
\begin{equation}
\begin{aligned}
V_{\textrm{t}}(r) =  k_BT ( V_{\textrm{r}}(r) + V_{\textrm{a}}(r))
\label{eq:Pot-total}
\end{aligned}
\end{equation}
\noindent where $V_{\textrm{r}}(r)$ is given by Eq.~\ref{eq:Yukawa} and $V_{\textrm{a}}(r)$ by a power law of the form
\begin{equation}
\begin{aligned}
V_{\textrm{a}}(r) =  - \epsilon_a \bigg( \frac{\sigma_{hs}}{r}\bigg)^\alpha
\label{eq:Pot-att}
\end{aligned}
\end{equation}

\noindent Here, we use a value $\alpha = 90$, which results in a range of about $4\%$ for the attractive contribution $V_{\textrm{a}}$, similar to what has been used previously for other globular proteins in order to reproduce their phase behavior and structural properties.

With this approach, we can now reproduce the measured data for both ionic strengths at all concentrations using a combination of $Z_{eff}^{RY} = 20$ and $\epsilon_a = 3.5 k_BT$, as shown in Fig. \ref{fig:mAb-G-S0-T25}, quite well. The only systematic deviation that we observe happens for the highest concentrations at the higher ionic strength, where a larger value of $\epsilon_a \approx 3.8$ $k_BT$ would be required, which is not consistent with the low concentration data. The estimated contact value of - 3.5 $k_BT$ for the attraction is found to be quite comparable to what has been used previously for 
globular proteins \cite{Cardinaux2007, Cardinaux2011, Wolf2014, Kastelic2015}. It is also instructive to look at the actual effective pair potentials for different concentrations for both ionic strengths plotted in Fig. \ref{Potential}. While the overall shape of $V_{\textrm{t}}(r)$ is similar in all cases, we see significant  differences between ionic strengths. At low ionic strength, $V_{\textrm{t}}(r)$ is characterised by a long-range soft-screened Coulomb repulsion, an energy barrier at about $r/R_{hs} \approx 2.1$, and then an attractive well with a depth of around 2 $k_BT$ up to contact at $r/R_{hs} = 2.$, where the hard core repulsion sets in. 

Overall this turns out to be comparable to the potential used to reproduce cluster formation in lysozyme, where one observed that a monomer-cluster transition would occur at the same temperature for a barrier height of about 2.5 $k_BT$. In our case, this would correspond to concentrations around 70 - 80 mg/ml, i.e. concentrations where we start to see an upturn in the measured $\big\langle R_{h} \big\rangle_{\mathrm{app}}$ values observed in DLS experiments (Fig. \ref{fig:SLS-DLS_mAbG}). For 50 mM NaCl added, the situation is quite different, with the barrier being always below 1 $k_BT$, indicating that the antibodies are likely to self-assemble into small clusters already at low concentrations. This is in agreement with the fact that we already see measurable temperature dependence in DLS and SLS experiments for concentrations larger than about 10 mg/ml (see Fig. \ref{fig:SLS-DLS_mAbG}).

\subsection{An improved model including anisotropy}
\begin{figure}[h]
\includegraphics[width=0.7\linewidth]{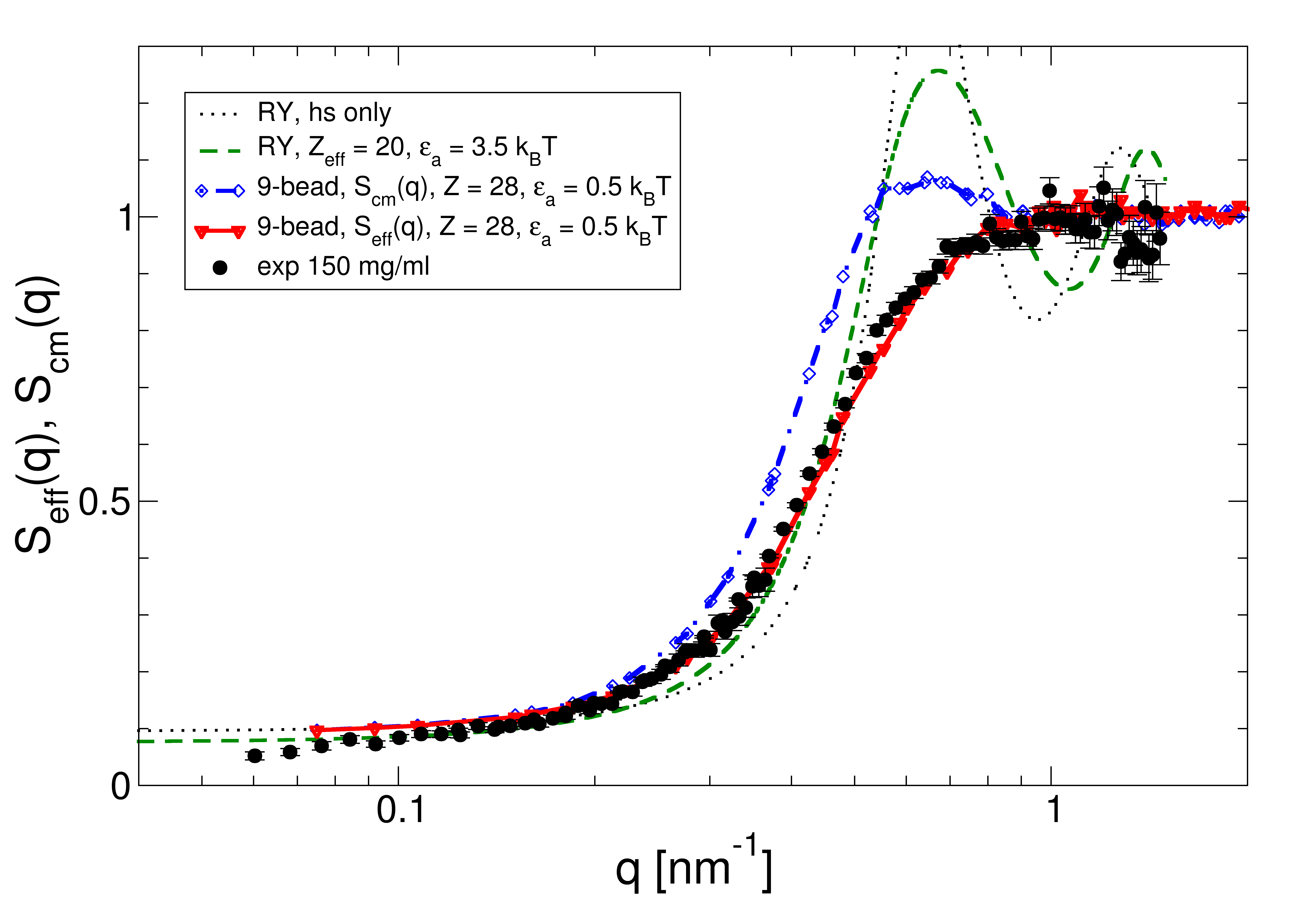}
\caption{Comparison between measured, calculated, and simulated effective structure factors for a simple colloid and a 9-bead hard Y model, respectively. The experimentally determined effective structure factor $S_{eff}(q)$ vs. \emph{q} is for $c = 150$ mg/ml and an ionic strength of 7 mM at 25 $^{\circ}$C (black-filled circles). The results obtained for a colloid model using an interaction potential based on screened Coulomb, excluded volume, and a short-range attraction as given by Eq. \ref{eq:Pot-total} based on the RY closure are given by the green dashed line (total charge $Z_{eff}^{RY} = 20$, hard-sphere diameter $\sigma_{hs} = 10$ nm, attraction strength $\epsilon_a = 3.5$ k$_B$T). The results from MC simulations using a 9-bead Y-model are shown as red triangles connected by the solid red line (bead diameter $\sigma_{bead} = 2.89$ nm, total charge $Z_{eff}^Y = 28$, attraction strength per bead $\epsilon_a = 0.5$ k$_B$T). Also shown are the results for a RY calculation using a hard sphere potential only (black dotted line), and the centre-of-mass structure factor $S_{cm}(q)$ obtained from the 9-bead simulation (blue dashed-dotted line).}
\label{Sq-150}
\end{figure}

While our simple colloid model of hard spheres interacting via a mixed effective pair potential described by Eq.~\ref{eq:Pot-total} is indeed able to reproduce the mesoscopic experimental structural quantity $\big\langle M_{w} \big\rangle_{\mathrm{app}}$, it obviously has important shortcomings. We have already commented that the effective charge $Z_{eff}^{RY} = 20$ needed to reproduce the measurements is too low compared to $ Z_{calc}$ obtained from the known molecular properties of the mAb, and also that the more microscopic structural correlations expressed by the measured effective structure factor $S_{eff}(q)$ are strongly overestimated at contact. This becomes even more obvious when looking at the comparison between the measured and calculated $S_{eff}(q)$ for a concentration of $c = 150$ mg/ml  at the low ionic strength of 7 mM shown in Fig. \ref{Sq-150}. While the osmotic compressibility, expressed by the asymptotic low-$q$ value $S(0)$, is well reproduced, the nearest neighbour peak predicted by the colloid model (green dashed line) is very pronounced, while completely absent in the measured data. We have already provided some qualitative arguments for the discrepancy between theoretical and measured structure factors linked to the mAb anisotropy in the preceding sections. In order to look more carefully into the reasons for this ultimate failure of the simple model, we have thus performed additional computer simulations on a less coarse-grained model that already contains anisotropic features mimicking the mAb structure more closely, while still allowing the investigation of highly concentrated systems with reasonable computational costs.

\begin{figure}[h]
\centering
\includegraphics[scale=0.6]{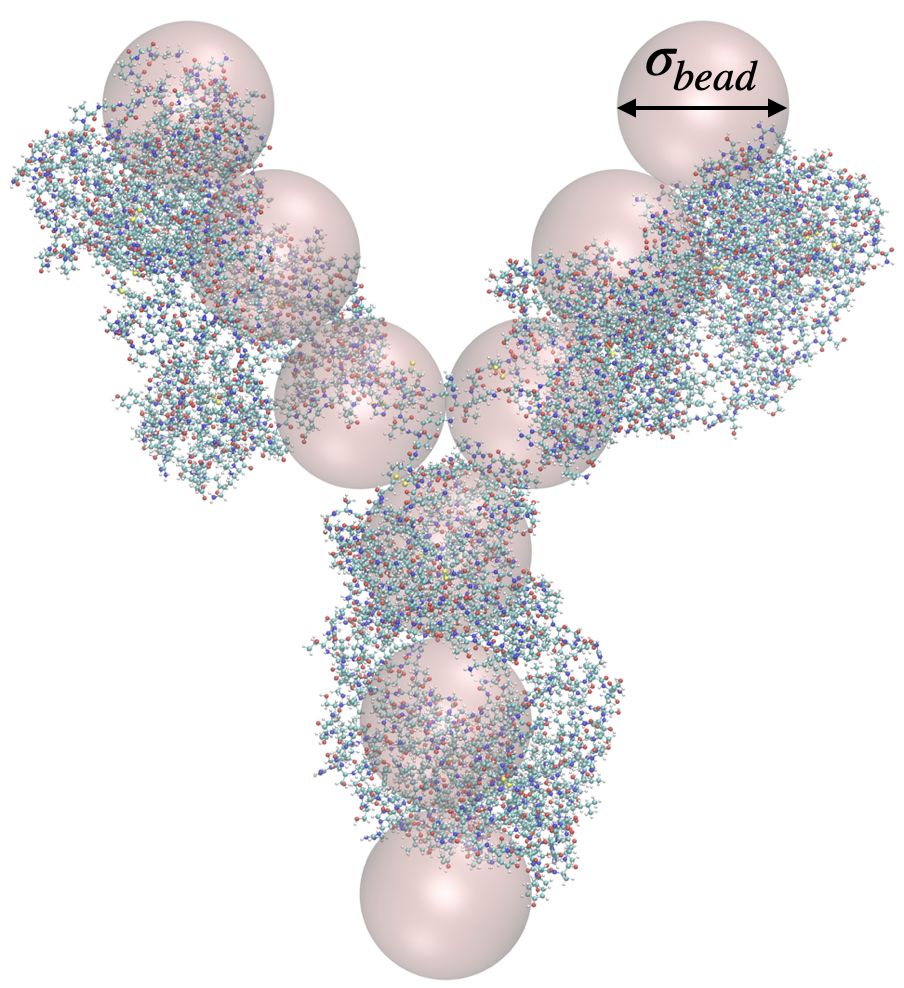}
\caption{9-bead Y model used in MC simulations of concentrated solutions, also shown is the all-atom structure. 
}
\label{9bead-model}
\end{figure}

The model follows a similar approach previously used by some of us in order to obtain insight into the self-assembly of mAbs at higher concentrations and is described in more details in the methods section \cite{Skar-Gislinge2019, Skar-Gislinge2023}. Each mAb consists of 9 beads arranged in a Y-shaped symmetric colloidal molecule, where each sphere has a unit-length diameter $\sigma_{bead}$, and where the radius of gyration of the 9-bead Y model is given by $R_g^{Y} = 1.7297 \sigma_{bead}$.
Each bead in the coarse-grained Y model is a hard sphere with diameter $\sigma_{bead}$ interacting with each other with infinite repulsive potential at contact, a screened Coulomb potential and an additional attractive contribution, similar to that used for the simple hard sphere model, given by 
\begin{equation} \label{potentialbead}
   V_b(r) = 
  L_{B} Z_{bead}^{2}  \left( \frac{e^{\kappa \sigma_{bead}/2}}{1 + \kappa \sigma_{bead} /2} \right)^{2} \frac{e^{-\kappa r}}{r} - \vphantom{ \left( \frac{e^{\kappa \sigma_{bead}/2}}{1 + \kappa\sigma_{bead} /2} \right)^{2}} \epsilon_a \left( \frac{\sigma_{bead}}{r} \right)^{6};  \text{ for r}  > \sigma_{bead} \\
\end{equation}
 \noindent and each antibody is treated as a rigid body. We assume that all beads are equally charged, with a charge $Z_{bead} = Z_{eff}/9$. 
 
 From the MC simulations, we can then calculate the effective structure factor for the 9-bead model using,
  \begin{equation}\label{Sqtot-sim}
\begin{split}
S^{eff}_{sim}(q) &= \frac{1}{P(q)} \frac{1}{9N}\left\langle \sum_{i_b=1}^{9N}\sum_{j_b=1}^{9N} e^{-i {\bf q} \cdot ({\bf r}_{i_b}-{\bf r}_{j_b})} \right\rangle \\
\end{split}
\end{equation}
 \noindent where the sum is taken over all beads of all antibodies, whose coordinates are $\textbf{r}_{i_b}, \textbf{r}_{j_b}$, and the average is taken over all trajectories.  Here, $P(q)$ is the form factor of a single 9-bead Y structure and we simulate $N = 1000$ hard Y molecules in the simulation. In order to compare results from simulations and experiments, the bead diameter is chosen in order to match the experimentally measured radius of gyration with the theoretical one, resulting in $\sigma_{bead} = 2.89$ nm.


\begin{figure}[h]
\includegraphics[width=0.8\linewidth]{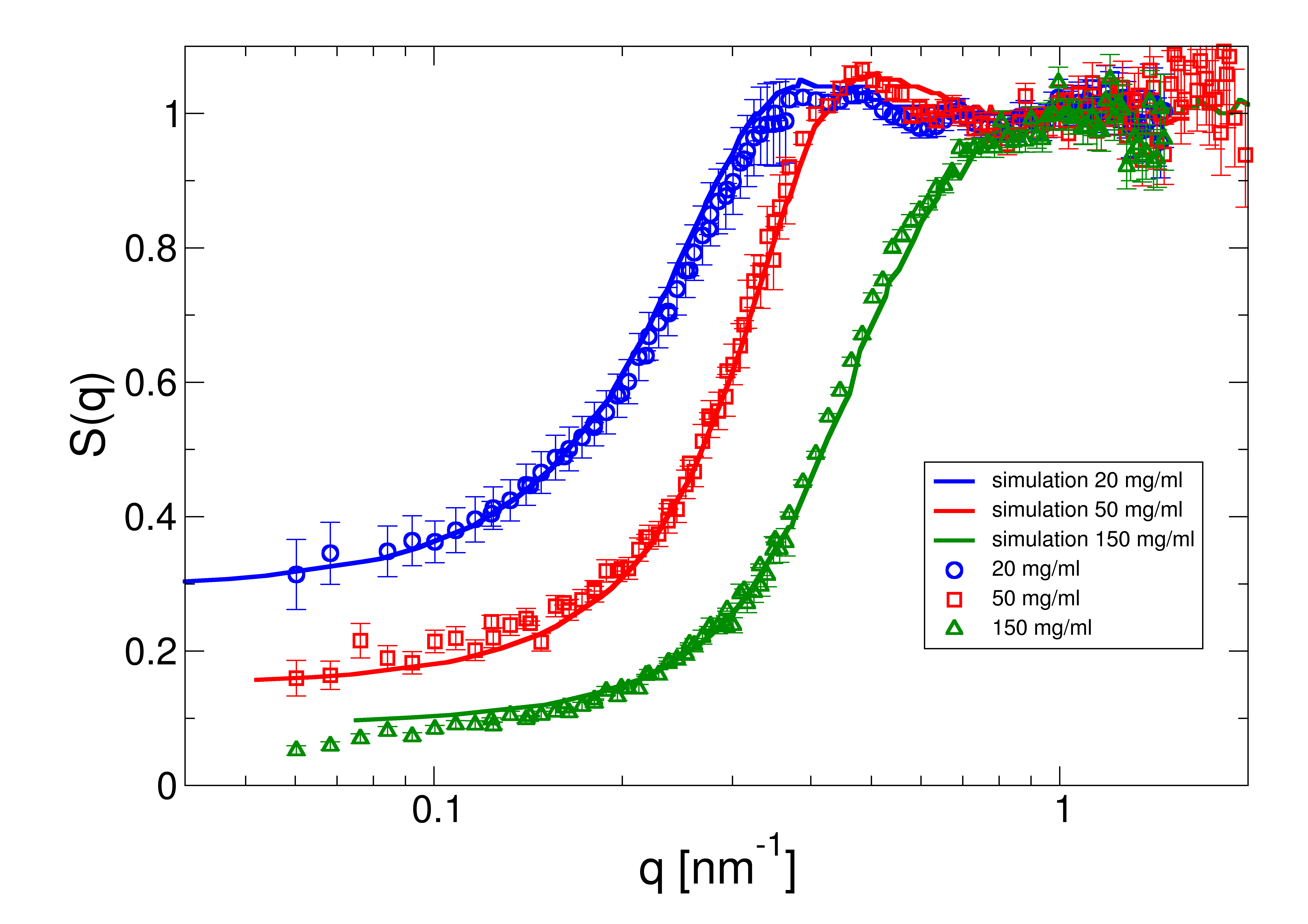}
\caption{Experimentally determined effective structure factor $S_{eff}(q)$ vs. \emph{q} for different concentrations and an ionic strength of 7 mM at 25 $^{\circ}$C compared to results from MC simulations using a 9-bead Y-model (bead diameter $\sigma_{bead} = 2.89$nm, total charge $Z_{eff} = 28$, attraction strength per bead $\epsilon_a = 0.5$ k$_B$T).  Experimental results for $c = 20$ mg/ml are shown as blue circles, $c = 50$ mg/ml as red squares and $c = 150$ mg/ml as green triangles.  Results from MC simulations using a 9-bead Y-model are shown as solid lines.
}
\label{Sq-conc}
\end{figure}

The results for the samples with no added salt are summarised in Fig. \ref{Sq-conc}. The agreement between measured and calculated effective structure factors is very good, in particular given the still very simple model and a high degree of coarse-graining. Furthermore, the total effective charge $Z_{eff}^{Y} = 28$ is now close to the theoretical one, calculated from the MC simulation using the molecular structure of the mAb. This clearly shows that while standard approaches using either electrophoretic light scattering, $B_2$ or $k_D$ measurements combined with the colloid models commonly used in the data analysis result in too low effective charges, 
SAXS combined with anisotropic bead models provides much more realistic values for the overall mAb charge.

We can now obtain further insight into the main reasons for the failure of the simple colloid model to correctly reproduce the true overall charge and the solution microstructure by also looking at the center of mass structure factor $S_{cm}(q)$ given by
\begin{equation}
\label{Sqcm-sim}
  S_{cm}(q) = \frac {1}{N}\left\langle 
 \sum_{i,j=1,N} e^{-i {\bf q} \cdot ({\bf r}_{i,cm}-{\bf r}_{j,cm})} \right\rangle,
\end{equation}
\noindent where  ${\bf r}_{i,cm}$ and ${\bf r}_{j,cm}$ are the coordinates of the centers of mass of $i$-th and $j$-th Y-molecule and $N$ is the total number of Y's in the simulation box, respectively. For monodisperse spherical particles, $S_{cm}(q) = S_{eff}(q)$, whereas this is not the case for anisotropic objects such as mAbs. Here, the total scattering intensity can no longer be described by independent contributions from particle shape (particle form factor $P(q)$) and interparticle correlation effects (structure factor $S_{cm}(q)$). In fact, for anisotropic objects the scattering intensity depends on the orientation of the particle, and for interacting particles the orientation between particle pairs at distances closer or smaller than their overall diameter is no longer uncorrelated or random. There have been attempts to overcome this problem and use approximate schemes such as the decoupling approximation given by 
\begin{equation}
\label{decoupling}
  S_{eff}(q) = 1 + \beta(q)[S_{cm}(q) - 1]
\end{equation}
\noindent where $\beta(q) = \left\langle \lvert F(\bf q) \rvert \right\rangle^2 / \left\langle \lvert F(\bf q) \rvert ^2 \right\rangle$ and $F(\bf q)$ is the orientation-dependent scattering amplitude of an anisotropic object \cite{Yearley2013, Corbett2017, Pedersen2001}. However, as shown previously, this approximation provides good results only for small $q$-values, and the comparison between the calculated structure factor from the colloid model and $S_{cm}(q)$ obtained with the 9-bead model shown in Fig. \ref{Sq-150} clearly demonstrates why. There are significant differences between $S_{eff}(q)$ calculated for the spherical colloid model using the RY closure and the potential given by Eqs.  \ref{eq:Yukawa}, \ref{eq:Pot-total}, and \ref{eq:Pot-att} and $S_{cm}(q)$ obtained from the MC simulations using the 9-bead model. While the low-$q$ limit given by $S(0)$ are almost identical in both cases, the structural correlations at shorter characteristic distances comparable with the nearest neighbor distance are much less pronounced for the anisotropic model than for the spherical colloid model, clearly demonstrating that the effective pair potential used for the colloid model is not a good approximation of the potential of mean force between Y-shaped anisotropic objects. As a result, one of the main ingredients of the decoupling approximation given by  Eq.~\ref{decoupling} is not working. Therefore, a successful use of a simple spherical colloid model would require a much softer potential than the hard-sphere one, acting at distances closer than the effective sphere diameter and a charge distribution that is not limited to the surface of the effective sphere.

\begin{figure}[tb]
\includegraphics[width=0.8\linewidth]{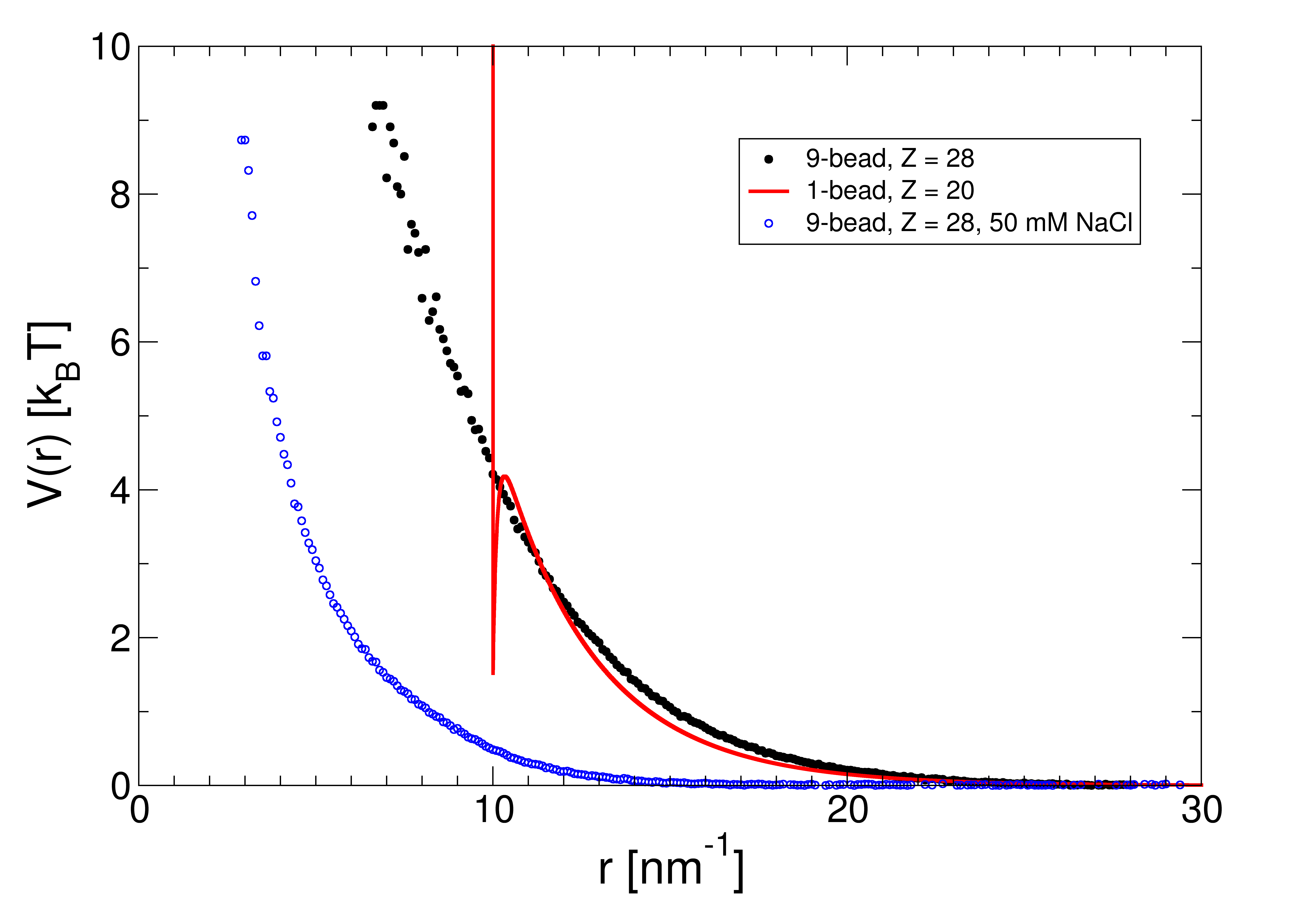}
\caption{Comparison between the theoretical pair potential as a function of the center-center distance $r$ as given by Eq. \ref{eq:Pot-total} with $Z_{eff} = 20$ and $R_{hs} = 5$ nm used for the colloid model (red solid line) and the effective potential of mean force PMF obtained from MC simulations using a 9-bead Y model (black dots), where the beads interact via Eq. \ref{potentialbead}, and with $\sigma_{bead} = 2.89$, total charge $Z_{eff} = 28$, attraction strength per bead $\epsilon_a = 0.5$ k$_B$T), respectively. Also shown is the PMF for the 9-bead Y model for an ionic strength corresponding to 50 mM NaCl added (blue open circles).}
\label{PMF}
\end{figure}

This is further illustrated in Fig.~\ref{PMF}, where we plot the interaction potential used for the colloid model as well as the effective potential of mean force (PMF) between the 9-bead Y particles, sampled from MC computer simulations (see Materials and Methods for details). At large distances, the screened Coulomb repulsion dominates in both cases and the two potentials overlap quite well up to a distance $r/R_{hs} \approx 1$. This is also the reason why the concentration dependence of the osmotic compressibility or $S(0)$ is reproduced well by both models, as it is primarily determined by the long-range repulsion except at extremely high concentrations or high ionic strength. At short distances, instead, the two potentials fundamentally differ, with the PMF for the 9-bead model continuing to increase up to much shorter distances. The comparison between the two potentials also directly shows the origin of the systematic differences between the effective charges obtained from the analysis of the experimental $S(q)$ data. At lower concentrations, where the experimental data ($S(q)$ and $S(0)$) are most sensitive to the value of $Z_{eff}$, $Z_{eff}$ is chosen such as to obtain a long-range potential that is capable of reproducing the measured data. Since in the colloid model all charges are distributed on the surface of the spherical particle with radius $R_{hs}$, the required charge is smaller than for a 3-dimensional charge distribution on the surface of a Y-shaped object such as a real mAb or the 9-bead model. 

It is thus clear that standard colloid models cannot be directly used  to infer the true overall charge of a mAb from experimental data such as measurements of the osmotic compressibility by static light scattering through a calculation of the second virial coefficient $B_2$ or from an analysis of the full structure factor $S(q)$ obtained from SAXS or SANS. The problem becomes even worse when using data such as the electrophoretic mobility or zeta potential or the interaction parameter $k_D$ from DLS measurements. While there exist attempts to calculate the zeta potential based on the molecular structure of mAbs, we currently lack the theoretical basis for performing scientifically correct calculations of the underlying electro-hydrodynamic problem for non-spherical objects with dimensions comparable to proteins. While such measurements thus provide information that is certainly interesting and helpful to estimate the overall colloidal stability of mAbs or obtain the charge sign, they cannot be used directly to quantitatively validate predictions for the overall charge and charge distribution based on the known molecular structure of a given mAb. In contrast, the use of a still highly coarse-grained model such as a 9-bead Y-shaped particle combined with SAXS measurements of the full structure factor results in much better estimates of the correct overall charge of a mAb.

However, there are additional important points that one needs to consider in any attempt to properly design an experimental study for a quantitative characterization of the overall net charge and the strength of an additional attraction of mAbs from SAXS. In addition to the bead diameter, which is chosen in order to match the mass distribution of the real mAb and the coarse-grained 9-bead Y as given by the radius of gyration, we have two free parameters, $Z_{eff}$ and $\epsilon_a$. It is thus important to look at how robust our choice for the values is when we analyze the measured $S(q)$ data. Therefore, we have conducted a systematic grid search procedure where we simulate mAb solutions with the 9-bead model at two ionic strengths and different concentrations for a large number of different values for the two free parameters $Z_{eff}$ and $\epsilon_a$. 
We then compared the experimentally measured and simulated effective structure factors and calculated the resulting overall error based on the so-called chi-square value given by 

\begin{equation}
\label{Chisquare}
  \chi^2 = \sum_{j=1,N} \frac{(S_{meas}^{eff}(q_j) - S_{sim}^{eff}(q_j))^2}{S_{sim}^{eff}(q_j)}
\end{equation}

\noindent where $S_{meas}^{eff}(q_j)$ is the measured and $S_{sim}^{eff}(q_j)$ the simulated value of the effective structure factor at a $q$-value $q_j$, and where the summation runs over all measured $q$-values. 

\begin{figure}[H]
	\includegraphics[width=1\linewidth]{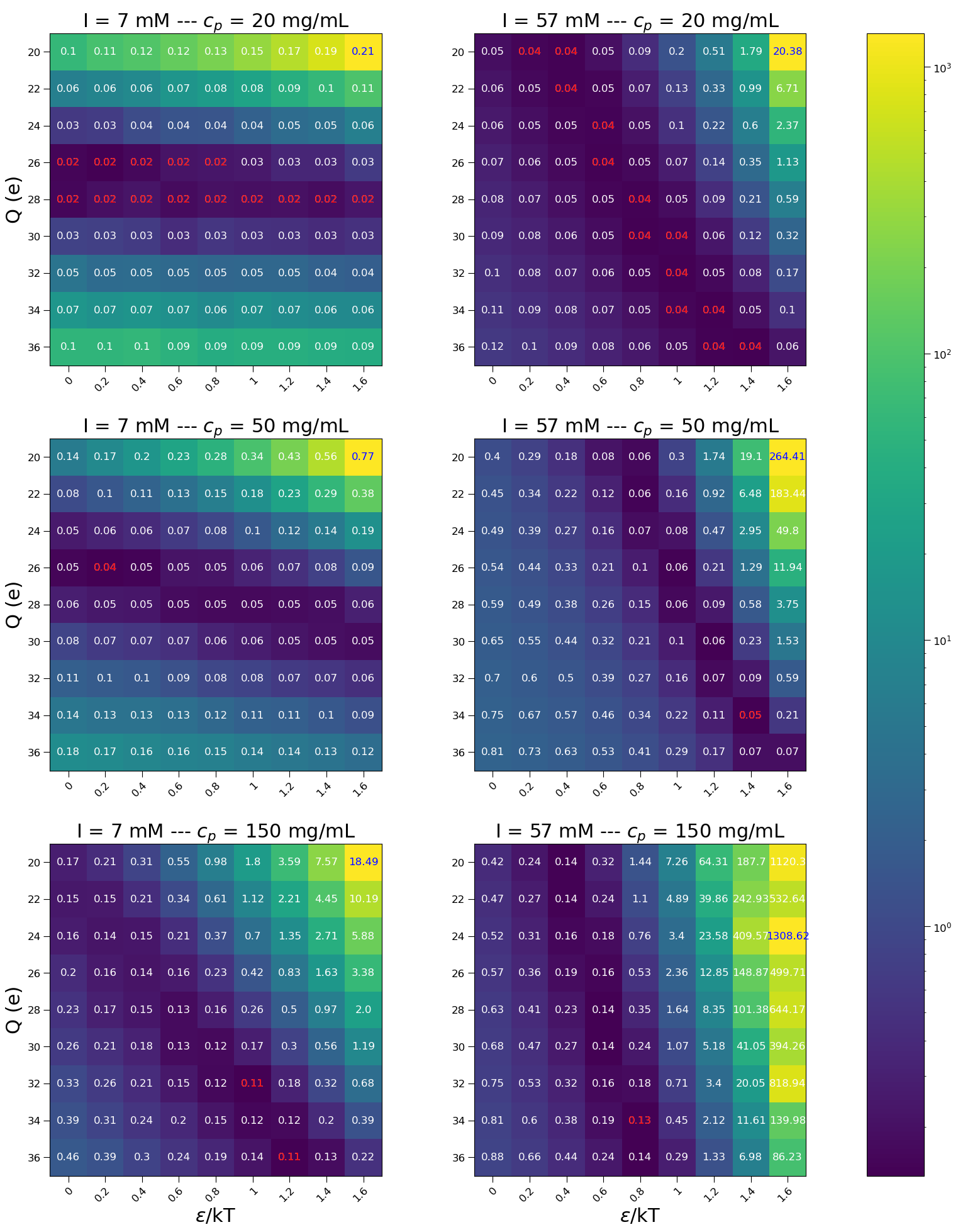}
	\caption{Resulting overall deviation between measured and simulated effective structure factors as given by $\chi^2$ defined in Eq. \ref{Chisquare} for three different concentrations and two ionic strengths as a function of the two parameters $Z_{eff}$ and $e_a$ in the bead potential given by Eq. \ref{potentialbead}. The color code used to describe the value of $\chi^2$ for a given set of parameters is shown on the right. }
	\label{chisquareplot}
\end{figure}

The results are summarized in Fig. \ref{chisquareplot}. Measurements at low ionic strength and relatively low concentrations around 20-30 mg/ml are ideal to determine $Z_{eff}$ with high accuracy from a single SAXS measurement for sufficiently charged mAbs. Under these conditions, the structural correlations are completely dominated by the long-range Yukawa contribution, and the influence of excluded volume and short-range attractions are negligible.

The nearest neighbor peak is thus most pronounced and its position depends entirely on the particle number density. On the other hand, the resulting $S_{sim}^{eff}(q)$ is insensitive to the choice of $\epsilon_a$ under these conditions. At higher ionic strength, the two parameters are now strongly coupled, and it is not possible to obtain accurate values from SAXS measurements at a single concentration. At high concentrations and low ionic strength, both parameters are also strongly coupled, and there is no unique parameter choice based on the $\chi^2$-evaluation only. Finally, at high ionic strength and high concentration, we obtain a more robust estimate of $\epsilon_a$, but the data is insensitive to the choice of $Z_{eff}$.


\subsection{Dynamic properties - DLS}

Having been able to reproduce the structural properties of the mAb solutions at both ionic strengths, we next proceed with an analysis of the experimentally observed concentration dependence of the collective diffusion coefficient or apparent hydrodynamic radius $R_{h,app}$ shown in Fig. \ref{fig:SLS-DLS_mAbG}. We use the same model of monodisperse spheres with a potential given by Eqs.  \ref{eq:Yukawa}, \ref{eq:Pot-total}, and \ref{eq:Pot-att}. We then follow the approach described by Neal et al. \cite{Neal1984} in the investigation of the structural and dynamic properties of Bovine Serum Albumine (BSA) at low ionic strength. The calculation of the short-time collective diffusion coefficient, $D_c^s(q)$, relies on pairwise additive hydrodynamic interactions, which should be accurate up to volume fractions of around $\phi \leq 0.05$. For our coarse-grained mAb model this roughly corresponds to $c \leq 25$ mg/ml.

In the calculation, we use the relationship between $D_c^s(q)$ and the ideal diffusion coefficient $D_0$, which takes place in the absence of interactions, given by \cite{Naegele1996, Banchio2008}
 \begin{equation}
	\label{Dcoll}
	D_c^s(q) = D_0 \frac{H(q)}{S(q)}
\end{equation}
\noindent where $H(q)$ is the hydrodynamic function that describes the effects of hydrodynamic interactions. We again use RY closure to calculate $S(q)$ and $g(r)$, while $H(q)$ is calculated as \cite{Neal1984}
\begin{equation}
	\label{Hq}
	H(q) = 1 + 6 \pi \rho R_h  \int_{0}^{\infty} r(g(r) - 1) \times \bigg[ \frac{\sin qr}{qr} +  \frac{\cos qr}{(qr)^2} -  \frac{\sin qr}{(qr)^3}\bigg] \,dr\ .
\end{equation}

\begin{figure}[!ht]
\centerline{\includegraphics[width=0.7\linewidth]{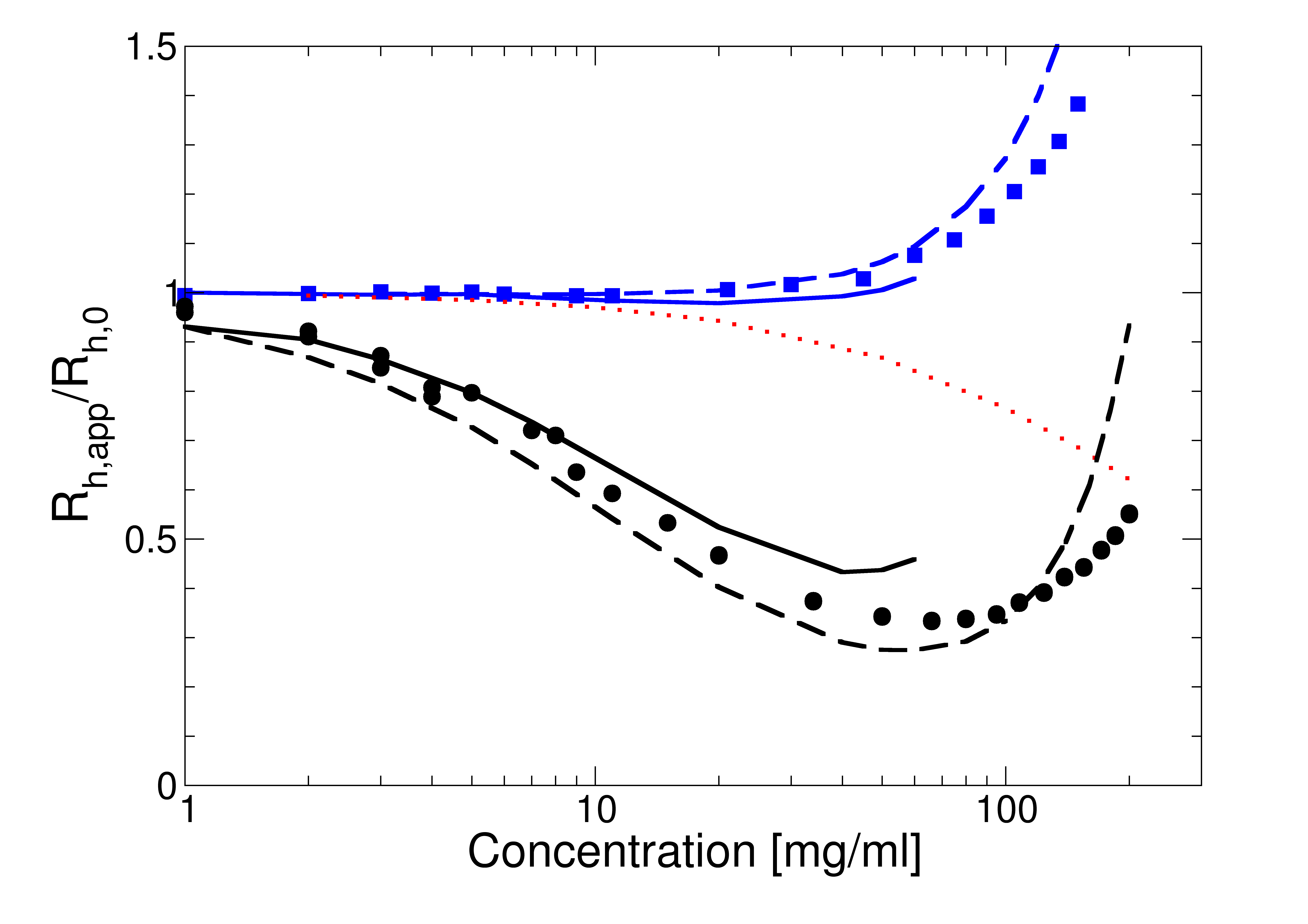}}
\caption{$R_{h,app}/R_{h,0}$ vs. \emph{c} compared to predictions using different colloid models for 25 $^{\circ}$C with no added salt (black symbols) and 50 mM NaCl added (blue symbols), where the solid lines are the predictions for $Z_{eff}^{RY} = 20$ and $\epsilon_a = 3.5 k_BT$ following the approach by Neal et al.\cite{Neal1984}. The dashed black and blue lines correspond to an ad-hoc description $R_{h,app}/R_{h,0} = S_{RY}(0)/H_{HS}(0)$, where $S_{RY}(0)$ is based on integral equation theory using the RY closure and $H_{HS}(0)$ is for pure hard spheres, respectively. Also shown is the theoretical result for hard spheres as the red dotted line.  }
\label{fig:mAb-G-Rhnorm}
\end{figure}

For small particles such as proteins, the measured diffusion coefficient corresponds to the so-called gradient diffusion coefficient given by 
\begin{equation}
	\label{Dc}
	D_c = \lim_{q \to 0} D_c^s(q) = D_0 \frac{H(0)}{S(0)} 
\end{equation}

\noindent where $H(0)$ is related to the sedimentation velocity, $U_{sed}$. In order to compare the DLS results with the calculated values, we therefore determine the asymptotic low-$q$ values $S(0)$ and $H(0)$ for the model parameters used to generate the data in Fig. \ref{fig:mAb-G-S0-T25}. The corresponding values $R_{h,app}/R_{h,0} = D_0/D_c$ vs. \emph{c} at $T$ = 25 $^{\circ}$C are shown in Fig. \ref{fig:mAb-G-Rhnorm} as the black solid line for no added salt and the blue solid line for 50 mM NaCl added, respectively. As a comparison, we also show the theoretical values for pure hard spheres \cite{Banchio2008}. The theoretical model for charged and weakly attractive spheres reproduces the experimental data for both ionic strengths surprisingly well, given the relatively simple underlying model that does not take into account the shape anisotropy and flexibility of the mAb. This shows that the coarse-grained short-range attractive and charged sphere model is not only suitable to  calculate thermodynamic and structural parameters such as the osmotic compressibility or $S(0)$ as well as local structure details such as the full static structure factor $S(q)$ at not too high concentrations, but also it allows us to estimate hydrodynamic interactions, characterised by $H(q)$, up to moderate concentrations.

However, despite the simplicity of the underlying model the calculations needed to obtain the theoretical values of $R_{h,app}/R_{h,0}$ vs. \emph{c} are still quite involved. In a recent DLS study of different mAbs, Dear et al. noted that their experimentally obtained values of $H(0)$ appeared to closely follow the theoretical predictions for hard spheres, irrespective of the specific solvent conditions and nature of the dominant protein interactions \cite{Dear2019}. We can therefore try a purely phenomenological approach in order to predict $D_c$ for our system, where we combine the theoretical RY values for $S(0)$ with $H_{hs}(0)$ for hard spheres in Eq.~\ref{Dc}. For $H_{hs}(0)$ we rely on the fact that $D_c$ follows a simple second-order virial expansion,  $D_c \approx D_0 (1 + k_D \phi)$, with $k_D = 1.45$, up to quite high concentrations $\phi \lesssim  0.3$ \cite{Banchio2008}. We can therefore calculate $H_{hs}(0)$ from this relationship combined with a calculation of $S(0)$ using the Carnahan and Starling approximation for the hard sphere $S_{hs}^{CS}(0)$: 
\begin{equation}
	\label{SCS}
	S_{hs}^{CS}(0) = \frac{(1-\phi)^4}{( 1+2\phi)^2 + \phi^3(\phi-4)}.
\end{equation}

\noindent This then results in the following approximation for $H_{hs}(0)$
\begin{equation}
	\label{HHS}
	H_{hs}(0) = (1 + 1.45 \phi) S_{hs}^{CS}(0),
\end{equation}
\noindent which allows us to estimate $R_{h,app}/R_{h,0}$ using Eq.~\ref{Dcoll} with a combination of a full RY calculation of $S(0)$ and $H_{hs}(0)$ from Eq.~\ref{HHS}. $R_{h,app}/R_{h,0}$ is then given by
\begin{equation}
	\label{Rhapp-RY-HS}
	R_{h,app}/R_{h,0} =  \frac{S_{RY}(0)}{H_{hs}(0)}
\end{equation}
\noindent where $S_{RY}(0)$ is the theoretical value of $S(0)$ calculated with RY, as shown in Fig. \ref{fig:mAb-G-S0-T25}. The corresponding values are also reported in Fig.~\ref{fig:mAb-G-Rhnorm}, 
and the observed agreement with experimental data is quite remarkable up to concentrations of about $c \lesssim  100$ mg/ml. However, this approach fails at predicting correctly the upturn in $R_{h,app}/R_{h,0}$  at higher concentrations, particularly for the lower ionic strength. 

\begin{figure}[!ht]
\centerline{\includegraphics[width=0.7\linewidth]{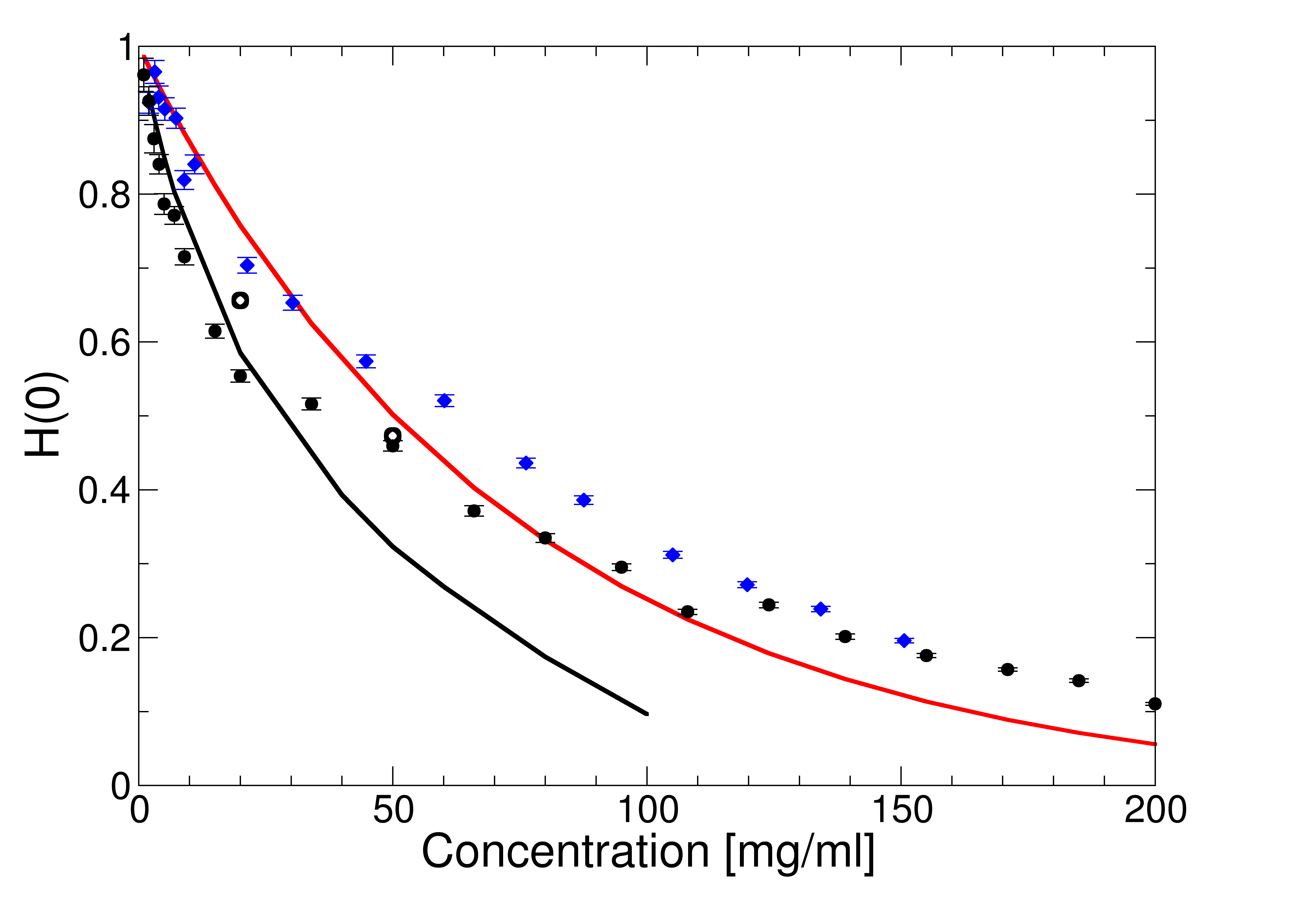}}
\caption{Experimentally determined hydrodynamic function $H(0) = S(0) / (R_{h,app}/R_{h,0})$ vs. \emph{c} compared to predictions using different colloid models for 25 $^{\circ}$C with no added salt (black solid circles) and for 50 mM NaCl added (blue solid diamonds), where the black solid lines are the predictions for $Z_{eff}^{RY}= 20$ and $\epsilon_a = 3.5 k_BT$ following the approach by Neal et al.\cite{Neal1984}. The red solid line corresponds to the hard sphere prediction $H_{HS}(0)$ given by Eq.~\ref{HHS}, and the two open black circles describe the results using Eq.~\ref{Hq} with the pair correlation function $g(r)$ obtained from the computer simulations with the 9-bead model, respectively. 
}
\label{fig:mAb-G-H0}
\end{figure}

However, the good agreement seen between the predictions of this ad-hoc model and the experimental data is somewhat misleading when using it as an argument that would support the hypothesis that the hydrodynamic function for mAb solutions is indeed well described by simple hard sphere theory. The surprisingly good agreement between prediction and experimental data for $R_{h,app}/R_{h,0}$ is partially caused by the small, but systematic overestimation of $S(0)$ when using the Rogers-Young closure, together with the chosen parameter values of $Z_{eff}^{RY} = 20$ and $\epsilon_a = 3.5 k_BT$ (see Fig. \ref{fig:mAb-G-S0-T25}). We can demonstrate this by directly comparing the experimentally determined values of $H_{exp}(0)$ with those calculated by using either Eq.~\ref{Hq} or \ref{HHS}, respectively, as shown in Fig. \ref{fig:mAb-G-H0}. Here, $H_{exp}(0)$  is obtained from
\begin{equation}
	\label{H0exp}
	H_{exp}(0) = S_{exp}(0) / (R_{h,app}/R_{h,0}),
\end{equation}
\noindent where $S_{exp}(0)$ is the experimentally measured $S(0)$ from SLS, $R_{h,app}$ the measured apparent hydrodynamic radius and $R_{h,0} = 5.4$ nm its asymptotic value for infinite dilution. Fig.  \ref{fig:mAb-G-H0} clearly shows that while the experimental data for the higher ionic strength at low concentrations $c \lesssim  50$ mg/ml are indeed well represented by hard sphere theory, this is not the case for the lower ionic strength. The calculation using Eq.~\ref{Hq},  together with the calculated pair correlation functions $g(r)$ from the Rogers-Young closure, on the other hand reproduces the experimentally measured hydrodynamic function quantitatively up to $c \lesssim  20$ mg/ml. When using the pair correlation function $g(r)$ obtained from the computer simulations of the 9-bead Y model instead of those for the sphere model, the experimental data is also accurately reproduced at $c = 50$ mg/ml (open black circles in Fig.  \ref{fig:mAb-G-H0}).

At high concentrations $c > 100$ mg/ml, where hard-core and attractive interactions become more important, the two sets of data for 0 mM and 50 mM NaCl approach each other. However, even under these conditions, the hard sphere approximation is not able to quantitatively reproduce the experimental data. For 50 mM NaCl, the error introduced by using the hard sphere model is approximately 40 $\%$ at $c = 100$ mg/ml and increases to 70 $\%$ at $c = 150$ mg/ml. Our data is thus somewhat at odds with the earlier findings in Dear et al. \cite{Dear2019}, although a closer look at their Fig. 3b also reveals systematic deviations between the measured and calculated $H(0)$ values at higher concentrations for one of their mAbs. While the simple hard sphere approximation combined with experimental SLS data thus allows us to make predictions that provide at least semi-quantitative trends for the concentration dependence of $R_{h,app}$ for mAbs, for highly charged molecules at low ionic strength such an approach also fails to reproduce the initial $c$-dependence, as quantified for example by the quantity $k_D$. Here, a more involved approach that takes into account a more quantitative description of the structural correlations and hydrodynamic interactions such as described by Eq.~\ref{Hq} is needed.

\subsection{Dynamic properties - viscosity}

In the previous section, we showed that the DLS and SLS data appear to be reasonably consistent when looking at them with the simple coarse-grained model of charged spheres with a weak short-range attraction. What remains unclear so far is whether the experimentally observed upturn in $R_{h,app}/R_{h,0}$ at high concentrations is also connected to the onset of self-association into equilibrium clusters under these conditions. While the theoretical approach used to calculate $R_{h,app}/R_{h,0}$ cannot be used at the highest concentrations where this upturn is quite prominent, the phenomenological model at least indicates that this could also be compatible with our simple colloid model and reflect the fact that short-time collective diffusion may also slow down at high concentrations, approaching an arrest transition \cite{Banchio2008, Foffi2014, Bucciarelli2016}. Clearly, our approach for calculating hydrodynamic properties is no longer accurate enough at high concentrations, where these observations are made. There are more advanced methods available, that have  been used for example to reproduce structural and dynamic properties of globular proteins, such as lysozyme, under conditions where they exhibit self-association into transient equilibrium clusters \cite{Das2018}. However, while they are theoretically and numerically much more involved than the approaches that we have used here, they are also not really quantitative under these conditions. Given that they also do not include possibilities to incorporate the strongly anisotropic shape and internal flexibility of the mAbs, we, therefore, abstain from using these models. Instead, we try to obtain more insight into possible self-assembly and cluster formation through a combination of phenomenological observations and their interpretation based on analogies to known systems with or without equilibrium cluster formation. Previous studies of cluster formation in protein solutions have clearly demonstrated that the relative viscosity $\eta_r = \eta_0 / \eta_s$, where $\eta_0$ is the zero shear viscosity of the antibody solution and $\eta_s$ the solvent viscosity, is a highly sensitive property that is strongly influenced by the formation of transient clusters \cite{Bergman2019}. We, therefore, take a closer look at the measured concentration dependence of $\eta_r$, and investigate whether our coarse-grained colloid model is able to reproduce the experimental data. Here we use the assumption that the strong increase of $\eta_r$ at high concentrations is primarily caused by excluded volume interactions as previously observed for various globular proteins \cite{Foffi2014, Bergman2019,  Skar-Gislinge2019, Skar-Gislinge2023}.

\begin{figure}[!hbt]
\centerline{\includegraphics[width=0.8\linewidth]{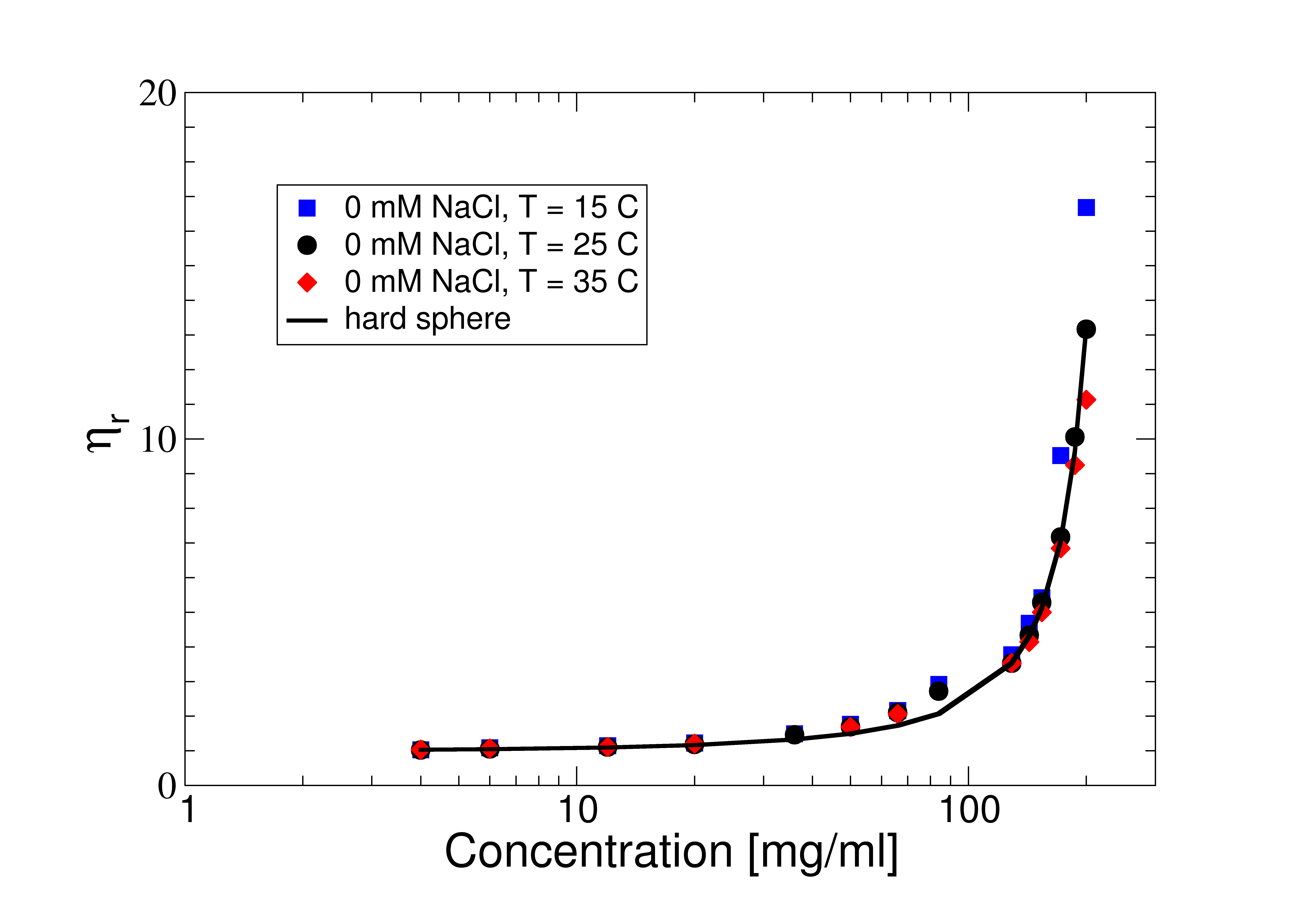}}
\caption{$\eta_r$ vs. \emph{c} for 15 $^{\circ}$C (blue symbols), 25 $^{\circ}$C (black symbols) and 35 $^{\circ}$C (red symbols),  with no added salt. Also shown is the theoretical result for hard spheres based on Eq. \ref{Quemada} as the black solid line. }
\label{fig:etar-mAbG}
\end{figure}

We make a first consistency test using the viscosity data obtained at low ionic strength, where, in the absence of well-defined charge patches with opposite signs, cluster formation should be negligible. The relative viscosity should thus be determined by the effective volume fraction of the monomers only. The experimental data at three temperatures (15, 25, and 35 $^{\circ}$C) is shown in Fig.  \ref{fig:etar-mAbG}. We can then compare this to the phenomenological Quemada expression frequently used for colloidal hard sphere systems\cite{Quemada1977}, 
 \begin{equation}
	\label{Quemada}
	\eta_r = \bigg(1 - \frac{\phi_{hs}}{\phi_{max}}\bigg)^{-2}
\end{equation}
where $\phi_{max}$ is the maximum packing fraction at which dynamical arrest occurs, which for hard spheres is around $\phi_{max} \approx 0.58$.

\noindent Using the experimental number densities and a hard sphere diameter $\sigma_{hs} = 10$ nm to calculate the effective hard sphere volume fraction $\phi_{hs}$, we thus obtain good agreement with the experimental values  up to concentrations around 150 mg/ml. This supports our choice of $\sigma_{hs}$, and also indicates that there is likely very limited self-assembly occurring under these conditions. At even higher concentrations, we find a visible temperature dependence of the relative viscosity, and we will need to come back to this point later when we discuss possible self-assembly at high concentrations in more detail.

\begin{figure}[!htb]
\centerline{\includegraphics[width=0.7\linewidth]{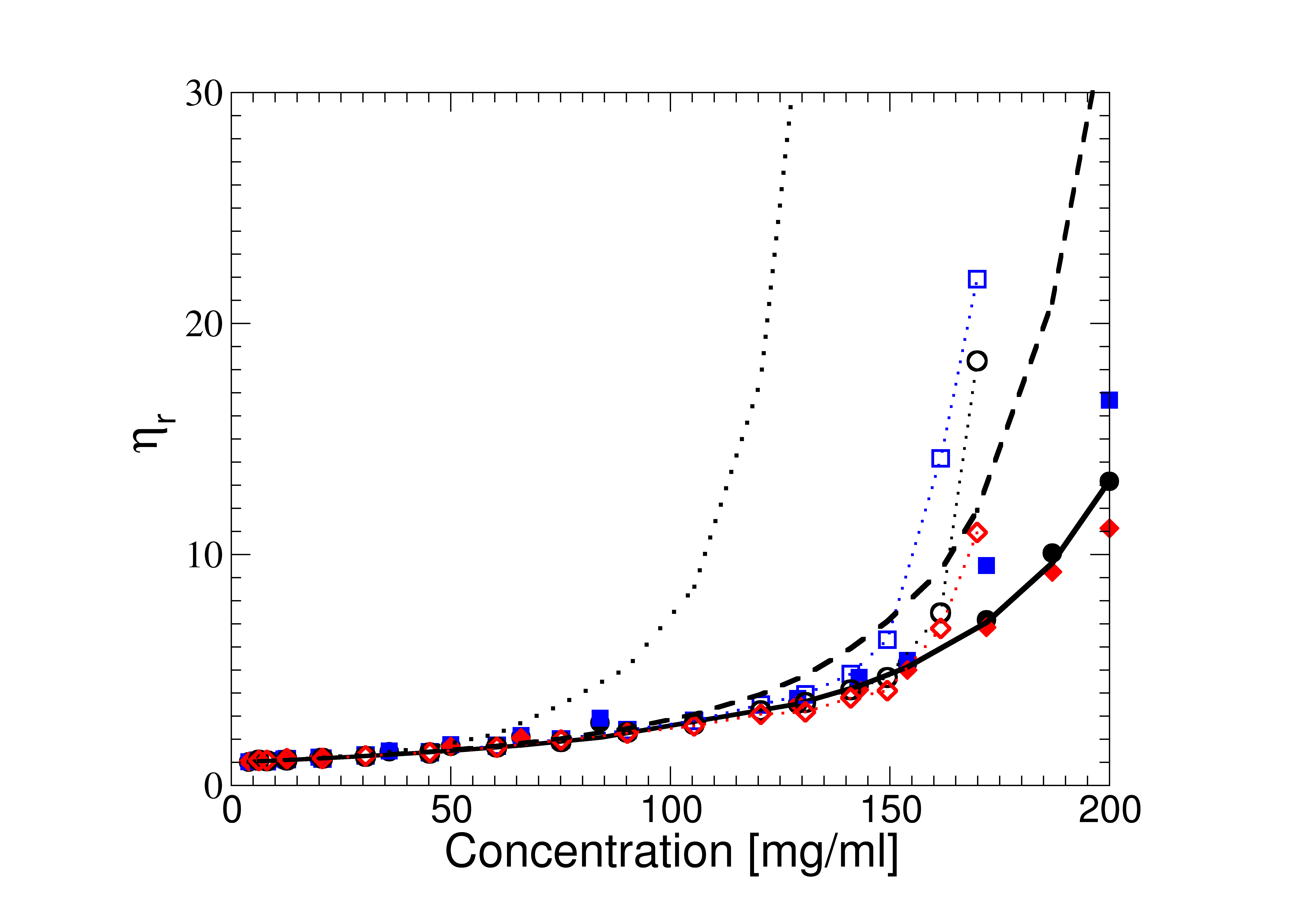}}
\caption{$\eta_r$ vs. \emph{c} for 15 $^{\circ}$C (filled blue symbols), 25 $^{\circ}$C (filled black symbols), and 35 $^{\circ}$C (filled red symbols),  with no added salt, and 15 $^{\circ}$C (open blue symbols), 25 $^{\circ}$C (open black symbols) and 35 $^{\circ}$C (open red symbols) with 50 mM NaCl added. Also shown is the theoretical result for hard spheres based on Eq. \ref{Quemada} with $\phi_{max} = 0.58$ as the black solid line and the prediction for weakly attractive hard spheres (Eq. \ref{Wagner}) as the black dashed line, respectively. The black dotted line shows the calculations for Eq. \ref{Wagner} using a concentration-dependent $\phi_{max}$ that follows the dependence upon $B_2^*$ given in ref. \cite{Eberle2012}.}
\label{fig:etar-mAbG-combined}
\end{figure}

We next consider the data at different ionic strengths and temperatures shown in Fig. \ref{fig:etar-mAbG-combined}, which reveals dramatic differences between the relative viscosity with no added salt and at a higher ionic strength with 50 mM NaCl added. Moreover, we see a clear temperature dependence at high concentrations that is much more pronounced at high ionic strength. While electrostatic interactions are known to influence suspension viscosity for charged colloids at low ionic strength, their effect should be much less pronounced for larger proteins with a relatively low effective charge at the ionic strength present with no added salt \cite{Heinen2012}. We would thus expect the relative viscosity to be slightly higher for no added salt but with a similar arrest transition in both cases. Weak attractive interactions are also known to influence the relative viscosity as well as the location of the arrest line in spherical colloids \cite{Krishnamurthy2005, Eberle2012}. A semi-empirical expression based on Eq. \ref{Quemada} has been derived in ref. \cite{Krishnamurthy2005} for a sticky sphere model and compared with data from colloidal systems. Here the relative viscosity is given by,
 \begin{equation}
	\label{Wagner}
	\eta_r = \eta_r^{hs}(\phi_{hs})  \bigg(1 + \frac{1.9\phi_{hs}^2}{\tau_b}\bigg)^{-2}
\end{equation}

\noindent where $ \eta_r^{hs}(\phi_{hs})$ is the relative viscosity of the pure hard sphere system given by Eq. \ref{Quemada}, and $\tau_b$ is the stickiness parameter that describes the strength of the attractive part of the potential. Eq. \ref{Wagner} is not restricted to a sticky hard sphere model, but can be used for arbitrary potentials $V(r)$ with weak attractions by matching the normalized second virial coefficient $B_2^* = B_2/B_2^{hs}$, where $B_2$ is the second virial coefficient defined as \cite{McQuarrie2000}

 \begin{equation}
	\label{B2}
	B_2 = 2 \pi  \int_{0}^{\infty}  \bigg(1 - e^{V(r)/k_BT}\bigg) r^2 \,dr\
\end{equation}

\noindent and $B_2^{hs} = 4(\pi \sigma_{hs}^3/6)$ is the second virial coefficient of the corresponding pure hard sphere system. The relationship between $B_2^*$ and $\tau_b$ is given by
 \begin{equation}
	\label{Tau-b}
	B_2^{*} = 1 -  \frac{1}{4 \tau_b} .
\end{equation}

In our case $B_2^*$ or $\tau_b$ are concentration dependent due to the changing screening conditions (Eqs. \ref{eq:Yukawa} and \ref{kappa}), and  thus we need to recalculate them for each sample condition. For the given sample and solvent conditions, and using the model of charged and weakly attractive spheres as described above, the $B_2^*$ values should decrease from 0.87 at the lowest concentration ($c = 3$ mg/ml) to 0.22 at the highest concentration ($c = 200$ mg/ml) considered in the calculations. The resulting theoretical curves, $\eta_r$ vs. $c$ for the samples with 50 mM added salt, are shown in Fig. \ref{fig:etar-mAbG-combined} together with the prediction for pure hard spheres and the experimental data for both ionic strengths. While we see from Fig. \ref{fig:etar-mAbG-combined}  that the weak attractions indeed should have a measurable effect on $\eta_r$, neither the magnitude nor the concentration dependence match the experimental observations for the data at higher ionic strength. When performing the calculation of $\eta_r$ using Eqs. \ref{Wagner} and \ref{Quemada}, we have to make an assumption for $\phi_{max}$, for which we have chosen $\phi_{max} = 0.58$. However, when looking at the available data for calculations based on similar model potentials used here \cite{Eberle2012}, we realize that for the values of $B_2^*$ found in this study for the higher ionic strengths, the corresponding values of $\phi_{max} $ would decrease from $\phi_{max} = 0.58$ at the lowest concentration to $\phi_{max} = 0.33$ at the highest concentration of $C = 169$ mg/ml, corresponding to a value of $B_2^* = 0.32$ based on the fits to the experimental SLS data. As a result, the corresponding calculated values of $\eta_r$ would exhibit a significantly stronger concentration dependence as also shown in Fig. \ref{fig:etar-mAbG-combined}, in fact, the system should then undergo an arrest transition at a concentration close to 165 mg/ml.

Fig. \ref{fig:etar-mAbG-combined} clearly indicates the limits of our highly coarse-grained approach of interpreting the experimental data for the mAb solutions using a model of weakly attractive charged hard spheres. While the structural and dynamic properties are well reproduced up to concentrations of about 50-100 mg/ml, and in the case of the osmotic compressibility or $S(0)$ even over the entire range of concentrations studied, the model is not capable of reproducing the dramatic increase of the relative viscosity at the highest protein concentrations at the higher ionic strength.

There is considerable evidence in the literature that the formation of (equilibrium) transient clusters can strongly influence the relative viscosity and for example result in a dynamic arrest through a so-called cluster glass transition, as long as the lifetime of the transient bonds between proteins or colloids is long enough \cite{Bergman2019, Cardinaux2007, Stradner2006b}. There are several possible mechanisms that can lead to self-assembly into equilibrium clusters. It is for example well documented that a combination of a long-range screened Coulomb repulsion and a short-range attraction can result in the formation of equilibrium clusters with a concentration-dependent size distribution \cite{Stradner2004, Stradner2006b, Cardinaux2011}. The presence of such clusters not only influences the measured values of $S(0)$ and $R_{h,app}/R_{h,0}$, but also the relative viscosity $\eta_r$, resulting in an arrest transition at lower concentrations when compared to a purely monomeric solution \cite{Bergman2019}. Other examples include cluster formation through attractive patches, either hydrophobic or charged patches of opposite sign such as often found in mAbs \cite{Skar-Gislinge2019}. There are in fact a number of studies where increased viscosity in concentrated solutions of mAbs is linked to cluster formation \cite{VonBulow2019, Yearley2014, Lilyestrom2013, Chowdhury2020, Buck2015, Godfrin2016, Skar-Gislinge2019}. Here we thus try to evaluate whether cluster formation as a source for the strong increase of the relative viscosity at high concentrations and ionic strength would be compatible also with the data from the static and dynamic scattering experiments. We follow a similar approach as already introduced in refs. \cite{Skar-Gislinge2019, Skar-Gislinge2023} to relate the average cluster size to the effective volume fraction and subsequently to the viscosity.

The starting point is the fact that the excluded volume of open or fractal clusters is larger than the excluded volume of the corresponding monomer solution. If we assume that clusters of size $s$ act as effective spheres with cluster radius $R_{cluster} \propto R_1 s^{1/d_F}$, where $R_1$ is the radius of a monomer, the effective cluster hard sphere volume fraction can be expressed as 

\begin{equation}\label{phiHS}
\phi_{hs, cluster} = \phi \langle N_{agg} \rangle_n^{(3 - d_F)/d_F},
\end{equation} 

\noindent where $d_F \approx 2 - 2.5$ is the fractal dimension of the clusters, $\phi$ is the nominal antibody volume fraction given by Eq. \ref{effhsvolume}, and $\langle N_{agg} \rangle_n = \sum n(s) s/\sum n(s)$ is the number average aggregation number given by the cluster size distribution $n(s)$ of clusters with size s. If we then assume that the relative viscosity of the cluster fluid is still given by Eq. \ref{Quemada}, but now with $\phi_{hs, cluster}$ instead of $\phi_{hs}$, the difference between the measured $\eta_r(c)$ and the calculated $\eta_r^{hs}(\phi_{hs}(c))$ provides us with an estimate of $\phi_{hs, cluster}$, which in turn allows the calculation of the average aggregation number $\langle N_{agg} \rangle_n$ through Eq. \ref{phiHS}. When we apply this approach to the two highest concentrations with 50 mM NaCl added, where we have observed a dramatic increase of the reduced viscosity (Fig. \ref{fig:etar-mAbG-combined}), we obtain values of $\langle N_{agg} \rangle_n = 1.5$ at $c = 160$ mg/ml and $\langle N_{agg} \rangle_n = 3$ at $c = 170$ mg/ml, respectively.

We then use an approach where clusters are treated as spherical particles with a hard sphere radius given by the hard sphere radius of the cluster $R_{hs}^{cluster} = R_{hs} \langle N_{agg} \rangle_n^{1/d_F}$, where $R_{hs} = 5$ nm is the hard sphere radius of the monomer, and a charge corresponding to $Z_{eff}^{cluster} = \langle N_{agg} \rangle_n Z_1$, where $Z_1 = 20$ is the effective charge of the monomer. The measured $S(0)$ or apparent aggregation number $\langle N_{agg} \rangle_{w,app}$ is then given by

\begin{equation}\label{naggapp_col}
\langle N_{agg} \rangle_{w,app} = \langle N_{agg} \rangle_w S^{eff}(0),
\end{equation}

\noindent where $\langle N_{agg} \rangle_w =  \sum n(s) s^2/\sum n(s) s$ is the true weight average cluster size and $S^{eff}(0)$ is the effective structure factor at $q = 0$ of a suspension of spheres with hard sphere radius $R_{hs}^{cluster}$, charge $Z_{eff}^{cluster} $ and number density $\rho_{cluster} = \rho / \langle N_{agg} \rangle_n$, interacting through a potential given by Eqs. \ref{eq:Pot-total} and \ref{eq:Pot-att}. In performing these calculations, we have to make assumptions for the polydispersity of the resulting cluster size distribution, for which we currently have no quantitative model. Therefore we have chosen values that correspond to the cluster size distribution of other self-assembling mAbs or globular proteins with the same average cluster size $\langle N_{agg} \rangle_n$ \cite{Cardinaux2007, Skar-Gislinge2019, Skar-Gislinge2023}. The resulting values are also given in Fig. \ref{fig:mAb-G-S0-T25} as the open blue circles. Given the very simple model and the number of assumptions made, the agreement is quite remarkable. We do expect that the model used will overestimate the charge effects as in the calculation we assume the charges to be spread on the surface of the effective hard sphere, whereas in the mAb cluster charges are also in the interior of the cluster, and screening thus starts not only on the surface.



\begin{figure}[!htb]
\centerline{\includegraphics[width=0.5\linewidth]{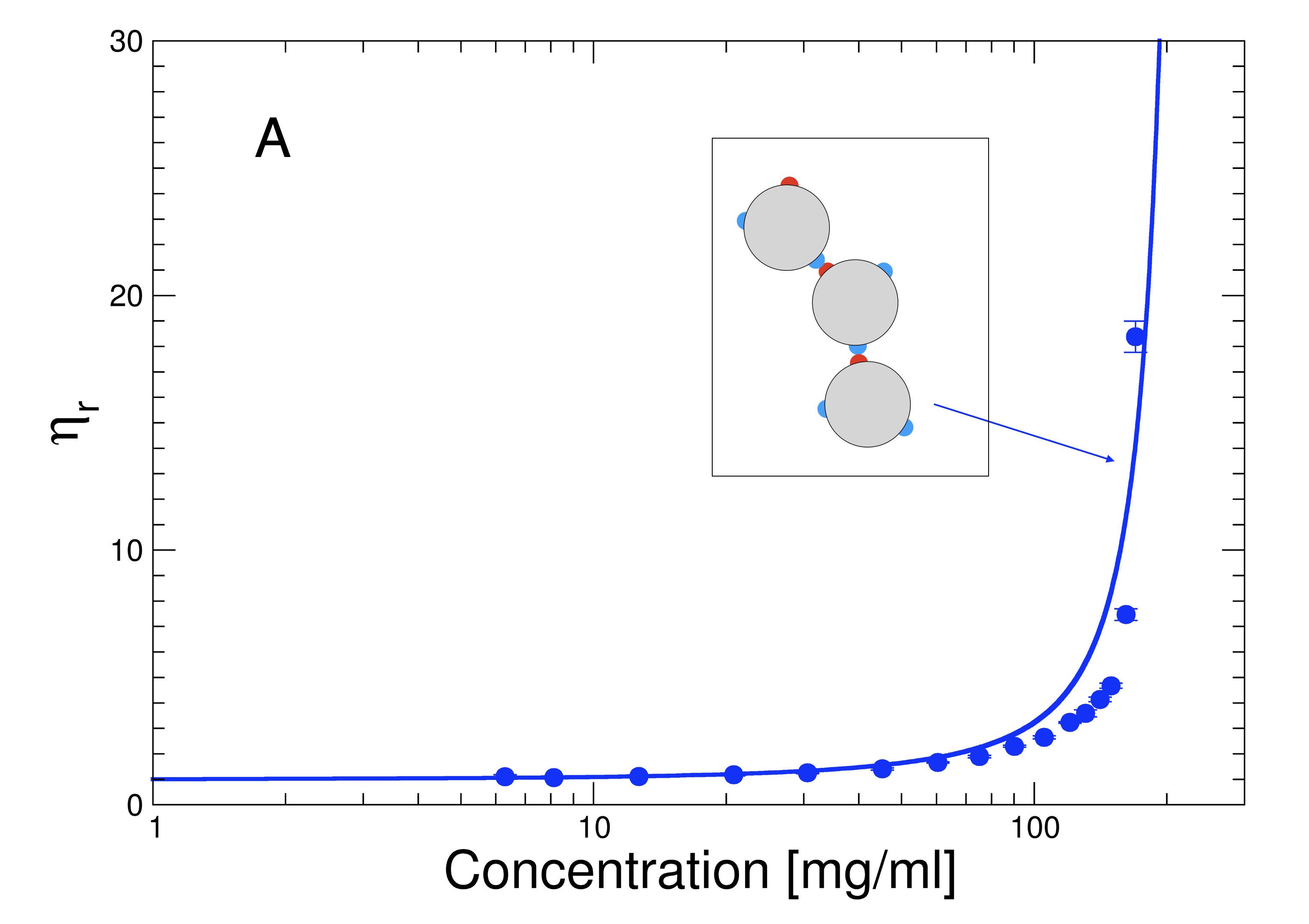}}
\centerline{\includegraphics[width=0.5\linewidth]{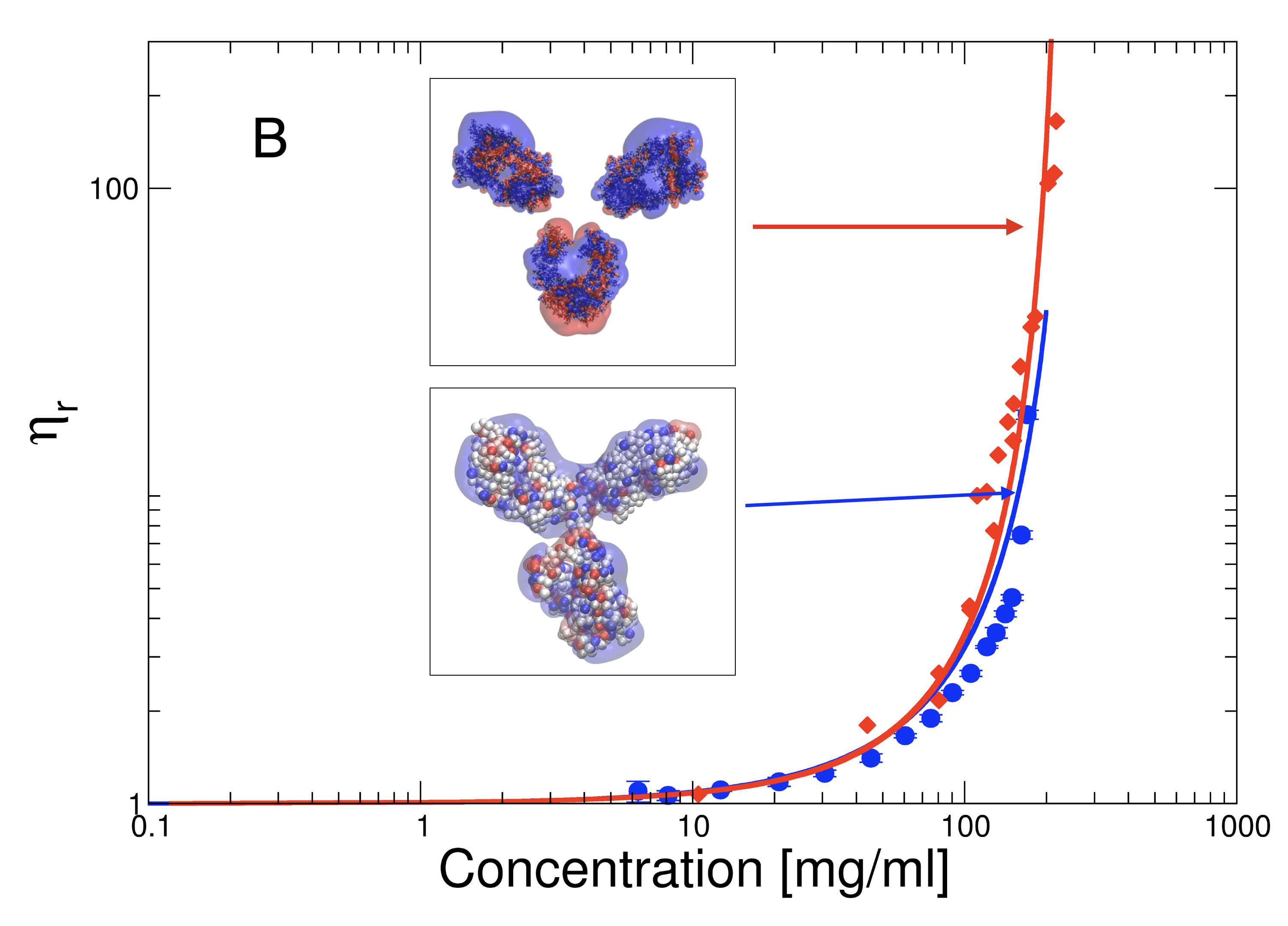}}
\caption{A: $\eta_r$ vs. \emph{c} for 25 $^{\circ}$C (filled blue symbols) with 50 mM NaCl added. Also shown is the theoretical result for a model of adhesive hard sphere clusters formed by patchy hard spheres (blue line), where the strength of the attraction between the patches is $\epsilon_{patch} = 7.14$ $k_BT$ (see refs. \cite{Skar-Gislinge2019, Skar-Gislinge2023} for details). B: Comparison between the relative viscosity of the current mAb (filled blue symbols) and the IgG4 mAb at 10 mM NaCl (red filled symbols, taken from ref. \cite{Skar-Gislinge2023}) described in ref. \cite{Skar-Gislinge2023}, together with the corresponding theoretical curves for the patchy sphere model (blue and red (taken from ref. \cite{Skar-Gislinge2023}) solid lines, respectively). Also shown as insets are the two isopotential surfaces of the two mAbs under these conditions. }
\label{cluster-mod}
\end{figure}

It is interesting to compare the viscosity data with previously published data on another mAb, an IgG4, where we have been able to demonstrate self-assembly into equilibrium clusters due to the interactions between well-defined patches of charges with opposite signs, and where the concentration dependence of the key experimental quantities such as $\eta_r$ could be quantitatively described by a simple coarse-grained model of hard spheres with attractive patches \cite{Skar-Gislinge2019, Skar-Gislinge2023}. Given the charge distribution on our mAb studied here described by the isopotential surfaces shown in Fig. \ref{CoarseGrainedModel}, where we observe a reasonably well-defined small negative patch on one end of the Fab region at 50 mM NaCl added, we can also attempt to use the same model of a patchy sphere as used in refs. \cite{Skar-Gislinge2019, Skar-Gislinge2023}, where we use a three-patch sphere model with one negative and 2 positive patches. Using the same patch size and range of the attractive patch-patch interactions as previously, the key model parameters are then the hard sphere diameter and the strength of the attractive square well potential. We use Wertheim theory in order to calculate the bond probability between attractive patches and thus the average aggregation number at each concentration and then a model of adhesive hard spheres to describe the interactions between the clusters and calculate $\eta_r$ as described in detail in Refs. \cite{Skar-Gislinge2019, Skar-Gislinge2023}. We use the same hard sphere diameter for our mAb as used to describe the structural and dynamic properties in the preceding sections ($\sigma_{hs} = 10$ nm), and adjust the attractive strength in order to obtain cluster sizes in agreement with the analysis of the viscosity at the highest concentrations measured, i.e. $\langle N_{agg} \rangle_n \approx 1.5$ at $c = 160$ mg/ml and $\langle N_{agg} \rangle_n \approx 3$ at $c = 170$ mg/ml, respectively. This results in $\epsilon_{patch} \approx 7.1$ $k_BT$, and the corresponding concentration dependence of $\eta_r$ for the adhesive hard sphere cluster model is shown in Fig. \ref{cluster-mod} together with the experimental data for 50 mM NaCl added.

With parameter values chosen to match the viscosity at the highest concentrations measured, it is obvious from Fig. \ref{cluster-mod}A that the simple model is not capable of reproducing the concentration dependence of $\eta_r$ even qualitatively for the mAb investigated in this study. While the model overestimates $\eta_r$ at intermediate concentrations $ c \leqslant $ 150 mg/ml, where the experimental data is in fact well described by the calculations for the presence of monomers only as shown in Fig. \ref{fig:etar-mAbG-combined}, it is also not capable to reproduce the steep increase of $\eta_r$ for $c >$ 150 mg/ml. This is quite in contrast to the data published previously for another mAb with well-defined charge patches, where the model reproduces the measured data almost quantitatively over the entire range of concentrations (see Fig. \ref{cluster-mod}B for a comparison). Clearly, a patchy sphere model with a concentration-independent attraction between patches is not able to reproduce our data. Instead, it looks as if self-assembly only sets in above a concentration of 150 mg/ml, also further supported by the fact that only then we observe a temperature dependence for $\eta_r$.  A comparison between the charge distribution and the resulting isopotential surfaces for both mAbs reveals some clear differences and provides further insight. For the IgG4 described in Ref. \cite{Skar-Gislinge2023}, we observe clear and well-defined patches of negative and positive charges on the Fab and the Fc regions, respectively, that also result in well-defined and clearly separated positive and negative isopotential surfaces. This allows for the formation of charge-driven equilibrium clusters at all ionic strengths, where self-assembly is in fact more pronounced at low compared to high ionic strength. Our current mAb however has a charge distribution where the significantly higher number of positive charges dominate and are well distributed over the surface of the mAb. At low ionic strength, this results in a positive isopotential surface that covers almost the entire mAb and only leaves a relatively small negative area at the tip of one of the Fab regions (Fig. \ref{CoarseGrainedModel}). While this would allow for a charge-driven temporary bond with a positively charged area, the long-range electrostatic interactions between the positive charges that are also illustrated by a significant overlap of the positive isopotential surfaces of two approaching mAbs result in a repulsive interaction at larger distances for all relative orientations. Therefore the formation of  long-lived clusters through transient bonds between regions of opposite charge that would influence the relative viscosity is unlikely to occur even at high concentrations. At high ionic strength, the situation is no longer so clear (see inset Fig. \ref{cluster-mod}B), and at high concentrations, where the additional contribution from dissociated counterions significantly contributes to the screening, such transient bonds may become possible above a threshold concentration. However, our highly coarse-grained model is not able to shed light on the underlying mechanisms relevant for the sudden strong increase of $\eta_r$ at the highest concentrations $c >$ 150 mg/ml. This will require much less coarse-grained models, where the actual charge distribution and other interactions between hydrophobic and/or hydrophilic residues are considered on a molecular level.

\section{Conclusions}

We have investigated the influence of antibody charge and solvent ionic strength on the structural and dynamic properties of dilute and concentrated mAb solutions using a simple and highly coarse-grained model of interacting colloids with spherical shape and a hard core. The interactions between mAbs can then be expressed by a centrosymmetric, effective pair potential composed of different contributions:  a hard core or excluded volume repulsion, a screened Coulomb interaction and a short-range attraction of unspecified origin, likely a combination of van der Waals and hydrophobic attraction. The model is found to be capable of reproducing the osmotic compressibility or apparent molecular weight of the mAb solutions over the entire concentration range investigated, i.e. up to weight concentrations as high as 200 mg/ml for no added salt and 170 mg/ml for 50 mM NaCl added. The only free parameters in the model are an effective hard sphere radius, an effective charge, and the strength of the attraction at contact. The values used for these quantities appear to be quite reasonable when judging from comparisons with previously published studies on globular proteins and their phase behaviour.

The model has not only allowed us to reproduce the static solution properties obtained by SLS, but also provided a consistent description of the concentration and ionic strength dependence of the collective diffusion coefficient measured by DLS, albeit up to lower concentrations of around 25 mg/ml only, due to the inherent limits of the theoretical approach used to calculate hydrodynamic interactions. Moreover, it has allowed us to correctly calculate the relative viscosity of the low ionic strength samples over the entire range of concentrations investigated.

However, our study has also clearly revealed the limits of the simple coarse-grained model, and pointed out areas where more work is needed. First of all, while it is capable of correctly reproducing the thermodynamic quantity apparent molecular weight at all concentrations for both solvent conditions, it only provides a correct description of the static structure factor as a measure of the local solution structure at low concentrations and for low ionic strength, where long-range weakly screened Coulomb repulsions dominate. At higher concentrations, where the overall screening length becomes shorter due to the counterion contributions, we see clear deviations between the measured and calculated S(q), indicating that the effective sphere model strongly overestimates local structural correlations. While this can be strongly improved by resorting to computer simulations based on a geometrical model that includes the anisotropy of the Y-shaped mAbs, it is not obvious how the shape and interaction anisotropy could be incorporated into a numerical model similar to the one used by us. One possible solution would be the use of a decoupling approximation \cite{Yearley2013, Corbett2017, Pedersen2001}. However, as demonstrated in Fig. \ref{PMF} such an approach is only promising if we also use a more appropriate model for the interaction potential that incorporates the softer potential of mean force and the more complex charge distribution experienced by mAbs. 

Moreover, the effective charge used in our calculations is a fit parameter that we cannot easily relate to the detailed molecular structure of the protein. While the simple colloidal model used here does allow us to reproduce the osmotic compressibility over the entire range of concentrations and ionic strengths investigated, and also correctly describes the collective diffusion coefficient over a reduced range of concentrations, it has no predictive power that would allow us to start from the molecular structure, estimate the total net charge and then calculate these experimental quantities. Furthermore, there are large differences between the effective charge obtained through electrophoretic light scattering and from the analysis of the static light scattering and SAXS data. In fact, the charge from ELS is found to be significantly smaller than that obtained from the structural or static data, which in turn is again lower than the theoretical charges $Z_{calc}$ obtained from state-of-the-art computer simulations using the molecular structure ($Z_{calc} \approx 31$ at low and $Z_{calc} \approx 36$  at high ionic strength, respectively)~\cite{Polimeni2023}. The fact that the charge from electrophoretic measurements is significantly smaller than the theoretical charge from the molecular structure is actually a common observation made in previous studies \cite{Yadav2011, Yadav2012}. The main problem here is that our data have been interpreted within a (consistent) model of a hard non-conducting sphere with a homogeneous surface charge. On the other hand, mAbs are Y-shaped anisotropic particles with a charge distribution that may not be homogeneously distributed on the exposed surface. While in our model screening of the effective charge starts at the surface of the hard sphere, in mAbs there are charges present inside the effective hard sphere radius, and screening starts at the position of the charge. This means that their contribution to the effective pair potential has already been reduced due to screening at a distance $R_{hs}$ away from the center of mass of the mAb, thus resulting in a lower effective charge $Z_{eff}$ for the hard-sphere model. Moreover, the friction coefficient of a mAb strongly depends on its orientation, and the electric field in an electrophoretic light scattering experiment together with the complex charge distribution and the corresponding presence of intrinsic and field-induced dipole moments can then lead to an orientation that may have a rather different friction coefficient than what is estimated from the hard-sphere model based on DLS experiments. While we can overcome some of these problems relating the molecular structure to the effective charge for the determination of $Z_{eff}$ from SLS or SAXS measurements using coarse-grained computer models, there is currently no theoretical basis to quantitatively calculate the electrophoretic mobility except through phenomenological approximations that lack truly predictive power. While $Z_{eff}$ obtained via electrophoretic measurements is clearly a valuable parameter that can be used to estimate solution stability and a propensity for self-assembly, the complexity of the underlying electrokinetic problem makes it unlikely that this will change soon.

Our work also provides guidelines for an efficient and precise determination of mAb charge from scattering experiments. Protein-protein interactions and in particular protein charge contributions are commonly determined from a series of measurements in the so-called virial regime at low concentrations, where the experimentally determined second virial coefficient $B_2$ or the diffusion interaction parameter $k_D$ can then be used to determine $Z_{eff}$ based on simple colloid models. Our data shown in Fig. \ref{chisquareplot} clearly show that, for reasonably charged mAbs, a single SAXS measurement at low ionic strength and low concentration combined with computer simulations of a strongly coarse-grained Y-shaped bead model results in highly accurate estimates of $Z_{eff}$, that are moreover reflecting the actual mean net charge of the protein. Such measurements can be obtained within a few seconds at a typical (bio)SAXS instrument at a synchrotron X-ray source, and within a few hours at a standard lab instrument, and require a minimum sample handling and amount of material. On the other hand, Figs. \ref{fig:mAb-G-S0-T25} and \ref{chisquareplot} also illustrate that attractive contributions are best obtained from a concentration series of SLS or SAXS measurements that includes high concentrations and also higher ionic strengths, where contributions from attractive interactions become more important.

Finally, our results also demonstrate that while coarse-grained models are able to reproduce all experimental quantities for the mAb investigated in our study at low ionic strength, they fail to predict the dramatic increase of the viscosity at high ionic strength and high concentrations. It is thus clear that we need a combination of simulations using less coarse-grained geometrical and interaction models in order to gain more insight. We could then try to develop strategies that would allow us to not only calculate effective charges based on the molecular mAb structure that could then be used with more refined models to calculate the most important structural and dynamic properties and their concentration, pH and ionic strength dependence, but also to define possible attractive patches that include contributions from hydrophobic or oppositely charged patches.

\section{Materials and Methods}
\subsection{Materials}
Experiments were performed with the monoclonal antibody Actemra (or Tocilizumab), an IgG1 that is an anti-IL-6 receptor. The samples used in this study were purchased commercially. Prior to experimentation, surfactant (polysorbate 80) was removed from the formulation using DetergentOUT Tween spin columns (G-Biosciences). Samples then underwent dialysis in 10,000 MWCO Slide-A-Lyzer cassettes (Thermo Fisher Scientific) to exchange into a basis buffer of 10 mM L-histidine at pH 6.0. Following buffer exchange, samples were concentrated to approximately 200 mg/mL using centrifugal concentrators (MilliPoreSigma). Samples were then filtered using 0.22 $\mu$m spin columns (Corning) and stored at -80 $^{\circ}$C prior to analysis.

Measurements were made with two buffer solutions at different ionic strength, \emph{i.e.}, 7 and 57 mM (equivalent NaCl). The H6 buffer corresponding to the buffer of the initially prepared stock solution was prepared by dissolving 5 mM of L-Histidine and 5 mM of Histidine-HCl Monohydrate (both Sigma-Aldrich, SE). The final pH of the buffer was adjusted to 6 $\pm$ 0.05 by the addition of a few microliters of hydrochloric acid (HCl, 0.1 M). With no added NaCl, this results in an ionic strength of 7 mM at the chosen pH = 6. For the H6 buffer with 57 mM ionic strength, 50 mM of NaCl  (Sigma-Aldrich, SE) was added to the original L-Histidine buffer, and the solution was then again titrated to pH 6 $\pm$ 0.05 with HCl, resulting in an overall ionic strength of 57 mM.

We adopted two different sample preparation protocols depending upon the ionic strength of the solution.
For the low ionic strength mAb solutions with an ionic strength of 7 mM, the samples at different concentrations were prepared by diluting the stock solution originally obtained with the low ionic strength buffer. The frozen stock solution was thawed at room temperature ($\approx$ 20$^{\circ}$C ) for $\approx$ 30 minutes, and then gently homogenized by using a micropipette. Once prepared, the samples were used for measurements, otherwise stored in a freezer at -80$^{\circ}$C. Before measurement, the concentration was measured via UV absorption spectroscopy, using a wavelength of $\lambda = 280$ nm and a specific absorption coefficient $\mathrm{E_{mAb, 1 \: cm}^{0.1\%, 280 \: nm}} =$ 1.51 $\mathrm{ml \cdot mg^{-1} \cdot cm}$. 
For the mAb samples prepared at 57 mM ionic strength, we used a different procedure. Here we exchanged the buffer of the stock solution using Amicon Ultra centrifugal filters of 10 kDa (Sigma-Aldrich, SE). The samples were centrifuged six times, and at each step, the buffer was removed and replaced with fresh one (H6, 57 mM ionic strength). The individual samples at different concentrations were then again prepared by diluting the high ionic strength stock solution with buffer of the same ionic strength (H6, 57 mM ionic strength), and the concentration was determined for each sample prior to the measurements as described above using the same extinction coefficient.
Freshly prepared samples were either used for measurements, or otherwise stored in a freezer at -85$^{\circ}$C.

\subsection{Dynamic and Static light scattering}
Dynamic (DLS) and static (SLS) light scattering measurements were performed with a goniometer light scattering setup (3D LS Spectrometer, LS Instruments, AG), implementing a modulated 3D cross-correlation scheme to suppress  multiple scattering contributions \cite{urban1998characterization,block2010modulated}, and with an ALV/DLS/SLS-5022F, CGF-8F-based compact goniometer system (ALVGmbH, Langen, Germany). The light source for the 3D LS Spectrometer is a 660 nm Cobolt laser with a maximum power of 100 mW, while for the ALV instrument it  is a Helium-Neon laser operating at a wavelength $\lambda$ of 632.8 nm with maximum output power of 22 mW. All measurements on the 3D LS Spectrometer were performed at a scattering angle $\theta =$ 110$^{\circ}$, corresponding to a scattering vector  $q =  (4\pi n/\lambda) sin(\theta/2) =$ 20.7  $\mu$m$^{-1}$, while those on the ALV instrument were performed at a scattering angle of $\theta =$ 104$^{\circ}$, corresponding to a scattering vector  $q =$ 20.8  $\mu$m$^{-1}$. Measurements were done at three different temperatures  (\emph{T}) of 15, 25, and 35 $^{\circ}$C. 
For DLS, intensity auto-correlation functions $g_{2}(q, \tilde{t})-1 $ vs. lag-time $\tilde{t}$ were analysed with a second-order cumulant function, using an iterative nonlinear fitting procedure \cite{frisken2001revisiting, mailer2015particle}:

\begin{equation}
    g_{2}(q, \tilde{t})-1 = B + \beta \Bigg\{ \mathrm{exp}{{(-\Gamma_{1} \tilde{t}}) \Big[ 1+\frac{1}{2}\mu_{2} \tilde{t}^{2} \Big] \Bigg\}^{2},}
\label{eq:1}
\end{equation}

\noindent where $B$ is the baseline, $\beta$ is the spatial coherence factor, $\Gamma_{1}$ is the relaxation rate (first cumulant) and $\mu_{2}$ is the second cumulant,  which characterizes deviations from the single exponential behavior. $\mu_{2}$ is related to the polydispersity of systems with $\sigma* = \sqrt{\mu_{2}}/\Gamma$, where $\sigma*$ is the normalized standard deviation of the size distribution. 
The apparent  hydrodynamic radius $\big\langle R_{h} \big\rangle_{\mathrm{app}}$ of the scattered object was then calculated via the Stokes-Einstein relation:

\begin{equation}
\big\langle R_{h} \big\rangle_{\mathrm{app}} = \frac{k_B \: T }{6 \: \pi \: \eta} \cdot \frac{q^{2}}{\Gamma_{1}},
\label{eq:2}
\end{equation}

\noindent where $\eta$ is the viscosity of the solvent at a given temperature and the term  $q^{2} /\Gamma_{1}$ is the inverse of the apparent collective diffusion coefficient $\big\langle D \big\rangle_{\mathrm{app}}^{-1}$.\\

For SLS, we calculated the so-called excess Rayleigh ratio ($\Re_{\mathrm{ex}}$) from the measured scattering intensity \cite{Schurtenberger1991}. For samples with no multiple scattering contributions, \emph{i.e.}, negligible turbidity:

\begin{equation}
\Re_{\mathrm{ex}} =\frac{I(q)_{\mathrm{sample}}}{I(q)_{\mathrm{toluene}}} \Bigg(\frac{n_{\mathrm{sol}}}{n_{\mathrm{toluene}}}\Bigg)^{2}\Re_{\mathrm{toluene}},
\label{eq:3}
\end{equation}

\noindent where $I(q)_{\mathrm{sample}}$ and $I(q)_{\mathrm{toluene}}$ are the scattered intensities of the sample and the reference solvent toluene, respectively; $n_{\mathrm{sol}}$ and $n_{\mathrm{toluene}}$ are the refractive indexes for the solvent and toluene; $\Re_{\mathrm{toluene}}$ is the Rayleigh ratio for toluene in cm$^{-1}$. For the 3D LS Spectrometer at $\lambda = 660$ nm and vertical/vertical polarized geometry (polarization of the incident and detected light) we have $\Re_{\mathrm{toluene}} = 0.8456 \times 10{-5}$ cm$^{-1}$, while for the ALV instrument with $\lambda = 632$ nm and vertical/unpolarized geometry we have $\Re_{\mathrm{toluene}} = 1.364 \times 10^{-5}$ cm$^{-1}$, respectively, at $T = 25$ $^{\circ}$C.\cite{Sivokhin2021}


 



Finally, the apparent molecular weight of mAb ($\big\langle M_{w} \big\rangle_{\mathrm{app}}$) as a function of concentration was then calculated using

\begin{equation}
\big\langle M_{w} \big\rangle_{\mathrm{app}} = \frac{\Re_{\mathrm{ex}}}{Kc}
\label{eq:5}
\end{equation}

\noindent where 

\begin{equation}
K = \frac{4\pi^2n^2_{\mathrm{sol}}\bigg(\frac{dn_{\mathrm{sam}}}{dc}\bigg)^2}{N_A \lambda_0^4}
\label{eq:6}
\end{equation}

\noindent $c$ is the mAb concentration in mg mL$^{-1}$, the ratio  $\frac{dn_{\mathrm{sam}}}{dc}$ is  the refractive index increment of the mAb (= 0.194 mL mg$^{-1}$), $N_{A}$ is the Avogadro number and $\lambda_0$ is the vacuum wavelength of the laser.

\subsection{Microrheology} 
Tracer particle microrheology experiments were performed via dynamic light scattering (DLS) in 3D cross-correlation mode as described in detail in refs. \cite{garting2018optical, garting2019synthesis}. Tracer particles were prepared according to ref. \cite{garting2019synthesis} using polystyrene particles (particle diameter \emph{d} = 300 nm) stabilized with covalently bonded 20 kDa poly(ethylene) glycol chains. 
For these measurements, a volume of 1 $\mu$L of the tracer particle stock solution was added to 100 $\mu$L of protein solution. The DLS measurements were carried out at a single scattering angle $\theta = $ 90$^{\circ}$ and at three different temperatures  (\emph{T}) of 15, 25, and 35 $^{\circ}$C. The addition of tracer particles in diluted or weakly scattering protein solutions results in a single-step relaxation process in the $g_{2}(q, \tilde{t})-1$ function, and the intensity auto-correlation functions were analysed with a first-order cumulant expansion \cite{frisken2001revisiting}: 

\begin{equation}
  g_{2}(q, \tilde{t})-1 = B + \beta \big\{  \mathrm{exp}{{(-\Gamma \tilde{t}})}\big\}^{2},
\label{eq:7}
\end{equation}

\noindent where $B$ is the baseline, $\Tilde{\beta}$ is the spatial coherence factor and $\Gamma$ is the relaxation rate.  The diffusivity of the tracer particle was then calculated as $D_{\mathrm{Sample}} = \Gamma q^{2}$. We then use the Stokes-Einstein relation to calculate the relative viscosity ($\eta_{r}$) through

\begin{equation}
  \eta_{r} = \frac{\eta_{\mathrm{\:Sample}}}{\eta_{\mathrm{0}}} = \frac{D_{\mathrm{\: Ref.}}}{D_{\mathrm{\: Sample}} }
\label{eq:8}
\end{equation}
 
\noindent where $\eta_{\mathrm{\:Sample}}$ and $\eta_{\mathrm{0}}$ are the solution and solvent viscosity, respectively, and $D_{\mathrm{\: Ref.}}$ refers to the diffusion coefficient of the tracer particles dispersed in the pure solvent.

\subsection{Small Angle X-Ray Scattering}

Small Angle X-Ray Scattering (SAXS) measurements were performed with a pinhole camera system (Ganesha 300 XL, SAXSLAB) equipped with a high brilliance microfocus sealed tube and  thermostated capillary stage. The accessible \emph{q}-range for these measurements was  from $5 \times 10^{-2} \lesssim q \lesssim 10$ nm$^{-1}$. Experiments were carried out at T = 15, 25, and 35 $^{\circ}$C. All measurements were corrected for the background radiation, buffer in the capillary, mAb concentration, and transmission, resulting in a normalized scattering intensity $\bigg(\frac{d\sigma}{d\Omega}(q)\;c^{-1}\bigg)$. The  experimental structure factors ($S(q)$), were calculated using

\begin{equation}
  S(q) = \frac{\Bigg[\frac{d\sigma}{d\Omega}(q)\cdot c^{-1}\Bigg]}{\Bigg[\frac{d\sigma}{d\Omega}(q)\cdot c_{0}^{-1}\Bigg]_{\mathrm{FF}}}
\label{eq:9}
\end{equation}

\noindent where $\bigg[\frac{d\sigma}{d\Omega}(q)\cdot c^{-1}\bigg]$ is the normalized scattered intensity at higher protein concentration $c$ and $\bigg[\frac{d\sigma}{d\Omega}(q)\cdot c_{0}^{-1}\bigg]_{\mathrm{FF}}$ is the normalized scattered intensity of the form factor at low mAb concentration $c_0$.

Additional synchrotron SAXS measurements were performed on beamline B21 at Diamond Light Source, Didcot, UK. The incident X-rays had a wavelength of 0.09524 nm (13 keV), with a sample-to-detector (EigerX 4 M) distance of 3.69 m, corresponding to a $q$-range of 0.045-3.4 nm$^-1$. Samples were loaded into the capillary using the BioSAXS sample robot. The temperature within the capillary and sample holder were set at T = 15, 25, and 35 $^{\circ}$C. The continuously flowing samples were exposed for at least 10 frames (depending on initial sample volume and concentration) where each frame corresponds to an exposure of 1 second. Prior to averaging, sequential frames were investigated for inconsistencies caused, for example, by the presence of radiation damage. This was achieved by both visual inspections of the frames and by fitting the Guinier region for each individual frame. Before and after each sample measurement, identical measurements were performed on the buffer. The buffer frames were averaged and subtracted from the sample scattering. Calculation of S(q) followed essentially the same procedure as used for the in-house SAXS, with 1 mg/ml data used as the form factor.


\subsection{Electrophoretic light scattering measurements}
The electrophoretic mobilities of the mAb samples were measured with a Zetaziser Nano ZS (Malvern Instruments Ltd., Malvern, U.K.) using DTS1070 folded capillary cells (Malvern instruments Ltd., Malvern U.K.). Stock solutions of mAb were prepared by dilution with buffer to reach the final mAb concentration (7-10 mg/ml), if required the ionic strength was adjusted by addition of NaCl to the dilution buffer. For each sample and temperature at least three repeat measurements were made. Prior to each measurement, the samples were left to equilibrate at the set temperature for at least 500 seconds.  The results for the electrophoretic mobility and the effective charge thus obtained are summarised in Table \ref{tab:mobility}

\begin{table}[h!]
\centering
\begin{tabular}{| c | c | c | c | c |} 
 \hline
   & 7 mM & 57 mM & 107 mM & 157 mM \\ [0.5ex] 
 \hline
 $\mu_e$ [$\times 10^{-4} cm^2/Vs$] & $0.6487 \pm 0.02 $& $0.3539 \pm 0.03 $&$ 0.2123 \pm 0.02 $& $0.1825 \pm 0.02$ \\ 
  \hline
 $Z_{eff}^\zeta$ & 12.8 & 13.4 & 10.1 & 10.0 \\ [1ex] 
 \hline
\end{tabular}
\caption{Results for electrophoretic mobility measurements at different ionic strengths at a concentration of 5 mg/ml at 7 mM ionic strength, and 6 mg/ml for the other ionic strengths.}
\label{tab:mobility}
\end{table}

The electrophoretic mobility $\mu_e$ of spherical particles is directly related to the effective charge $Z_{eff}^\zeta$ of the particle via \cite{Ware1974, Miller2020} 

 \begin{equation}
	\label{mobility}
	Z_{eff}^\zeta = \frac{\mu_e f}{e} \frac{1 + \kappa a_\zeta}{f'(\kappa a_\zeta)}
\end{equation}

\noindent where $f = 6 \pi \eta R_h$ is the hydrodynamic friction coefficient and $f'(\kappa a_\zeta)$ is a function that accounts for the electrostatic screening of the particle (or macroion) by the counterions. Here $a_\zeta = R_{hs} + R_{ci}$ is the particle radius including the Stern layer, where we use $R_{ci} = 0.18$ nm as the radius of the counterion. $f'(\kappa a_\zeta)$ is given by Henry's function \cite{Henry1931}, which we calculate using the form given by Swan and Furst \cite{Swan2012}. 


\subsection{Computer simulations}

We first calculate a representative solution structure of the mAb as the basis of the coarse-grained structure using homology modeling. The primary amino acid sequence was retrieved from patent US20120301460. A homology model was prepared using the Antibody Modeler module in Molecular Operating Environment (MOE) 2020 \cite{MOE}. Briefly, the primary sequence was used to identify suitable existing structures for the framework and variable domains upon which the model was built. The complementarity-determining regions (CDRs) were modelled individually based on known loop structures and were then grafted onto the antibody framework. The structure then underwent energy minimization using 'LowModeMD' to eliminate steric clashes.

Based on this structure, and using the same protocol as in Ref.~\cite{Mahapatra2022}, we construct a coarse-grained representation of the mAb by replacing each amino acid with a spherical bead of diameter $\sigma_{bead}^{aa} = (6M_{W,aa}/\pi \rho)^{1/3}$, where $M_{W,aa}$ is the amino acid molecular weight (in g mol$^{-1}$), $\rho = 1$ (in g mol$^{-1}\AA^{-3}$ ) is an average amino acid density \cite{Kaieda2014WeakSO}, and the suffix $aa$ stands for 'amino acid'. With the amino-acid-based coarse-grained model, we perform Metropolis-Hastings Monte Carlo (MC) simulations of the mAb solution using Faunus \cite{Stenqvist2013FaunusA}, which is a software allowing for several types of MC simulations, in order to estimate the mAb net charge $Z_{calc}$ and the charge distribution (as performed here \cite{Mahapatra2022}).

We then performed Monte Carlo (MC) simulations of the mAb solution reproducing the experimental conditions, such as the protein concentration, the solution pH, and the ionic strength, using bead models in Faunus. We adopted a coarse-grained 9-bead model for the mAb (see Figure \ref{9bead-model}), where each antibody consists of 9 beads arranged in a Y-shaped symmetric colloidal molecule, where each sphere has a unit-length diameter $\sigma$. The three central beads are arranged in an equilateral triangle, and the three arms of the Y, each made of three spheres, form angles of $150^{\circ}$ and $60^{\circ}$ with each other. The geometric construction of the antibody implies that the circle tangent to the external sphere has a diameter 
$d_Y \approx 6.16\sigma$. 
Each bead in the coarse-grained Y model is a hard sphere with infinite repulsive potential at contact and each antibody is treated as a rigid body. The individual beads interact in a continuum medium with a potential $V(r)$ described in Eq. \ref{potentialbead}. 

The solution properties were sampled by performing MC moves, such as molecule translation and rotation, on systems composed of 1000 mAbs in a cubic simulation box of 
side length $L$ needed for reproducing the experimental mAb concentration. The volume of the box was then calculated in the unit of Å$^{3}$ as $V = L^{3} = N_{p}M_{w}/(c_{p}N_{a}1e^{-27})$, where $M_{w}$ = 148 kDa is the mAb molecular weight, $N_{p} = 1000$ is the number of mAbs in the box, $c_{p}$ is the experimental mAb concentration in mg/mL, and $N_{a}$ is the Avogadro number.

We computed both the solution, or center of mass structure factor, $S_{cm}(q)$, taking into account the molecular mass centers as single points scatterers, and the effective structure factor, $S^{eff}(q)$. where each bead is consdiered as a single point scatterer. The center of mass structure factor is defined as, 
\begin{equation} 
    S_{cm}(q) = \frac{1}{N} \left\langle \sum_{i,j}^{1,N} 
    \exp^{i\vec{q}\cdot(\vec{r}_i-\vec{r}_j)}
    \right\rangle
\label{eq:Sq}    
\end{equation}
where $N$ is the number of the scatterers, i.e., the numbers of mAbs in the simulation, and $\vec{r}_i$ is the position vector of the $i$-th mAb. The average indicates an average over configurations and wavevector orientations. 
On the other hand, the second is calculated as, 
The effective structure factor is defined as:
\begin{equation} 
    S^{eff}(q) = \frac{S^{*}(q)}{P_Y(q)},
\label{eq:Sqmeas}    
\end{equation}
where $S^{*}(q)$ is still obtained from Eq.~\ref{eq:Sq}, but now considering each bead as a single point scatterer, and $P_Y(q)$ is the form factor of the 9-bead Y model. In both cases, the sampled \textit{q}-interval is $2\pi/L$, $2\pi p_{max}\sqrt{3}/L$, where \textit{L} is the box side length. 

We also used simulations in order to extract the so-called potential of mean force (PMF). Here, we perform simulations with two identical mAbs described by the 9-bead model shown in Fig.~\ref{9bead-model}, mAb-1 and mAb-2, which are aligned and placed at a given distance on the z-axis of the coordinate system of the simulations. During the simulation, mAb-1 can only rotate with respect to its centre of mass, while mAb-2 can also rigidly translate back and forward along z. The beads on the two mAbs interact through the potential given by Eq. \ref{potentialbead}. This then allows us to sample the PMF as a function of the centre of mass distance by using the flat histogram method \cite{Wang, ENGKVIST199663, Hunter}.

\section{Acknowledgments}

This work was financially supported by Sanofi and the Swedish Research Council (VR; Grant No. 2016-03301, 2018-04627, 2019-06075 and 2022-03142). We benefited from access to the Diamond Light Source, Didcot, UK, where part of the SAXS measurements was performed at the beamline B21, and has been supported by iNEXT-Discovery, project number 871037, funded by the Horizon 2020 program of the European Commission, and we gratefully acknowledge the help of the local contacts Nathan Cowieson, Katsuaki Inoue, and Nikul Khunti. The computer simulations were enabled by resources provided by the National Academic Infrastructure for Supercomputing in Sweden (NAISS) and the Swedish National Infrastructure for Computing (SNIC) at Lund University partially funded by the Swedish Research Council through grant agreements no. 2022-06725 and no. 2018-05973.

\bibliography{mAb-bib}

\end{document}